\documentstyle[12pt,aaspp4]{article}
\begin{document}
\title{Large Angular Scale Polarization 
of the
 Cosmic Microwave Background Radiation and the Feasibility of its Detection}

\author{Brian Keating, Peter Timbie}
\affil{University of Wisconsin, Madison}
\authoraddr{keating@wisp5.physics.wisc.edu, timbie@wisp5.physics.wisc.edu 
Department of Physics 1150 University Ave. Madison, WI 53706}

\author{Alexander Polnarev}
\affil{Queen Mary and Westfield College}
\authoraddr{Astronomy
 Unit, School of Mathematics, Mile End Rd., London UK E14NS}
\and
\author{Julia Steinberger}
\affil{Brown University}
\authoraddr{ js@het.brown.edu 
 Department of Physics Box 1843 Providence, RI 02912 }
\newcommand{\lan}{\langle}
\newcommand{\ran}{\rangle}
\newcommand{\be}{\begin{equation}}
\newcommand{\ee}{\end{equation}}

\begin{abstract}
In addition to its spectrum and temperature anisotropy, the 2.7 K Cosmic
Microwave Background (CMB) is also expected to exhibit a low level of
polarization.  The spatial power spectrum of the polarization can provide
details about the formation of structure in the universe as well as its
ionization history.  Here we calculate the magnitude of the CMB polarization in
various cosmological scenarios, with both an analytic and a numerical method.
We then outline the fundamental challenges to measuring these signals and focus
on two of them: achieving adequate sensitivity and removing contamination from
foreground sources.  We describe the design of a ground-based instrument (POLAR)
that could detect polarization of the CMB at large angular scales in the next
few years.
\end{abstract}

\section{Introduction}

\label{secintro}

The 2.7K Cosmic Microwave Background (CMB) radiation is a vital probe of all
modern cosmological theories.  This radiation provides a ``snapshot'' of the
epoch at which radiation and matter decoupled, approximately 300,000 years after
the Big Bang, and carries the imprint of the ionization history of the
universe. This information tightly constrains theories of cosmological structure
formation.

The three defining characteristics of this radiation are: its spectrum, spatial
anisotropy, and polarization.  The spectrum and anisotropy of the CMB have both
been extensively studied.  The {\em COBE} FIRAS has determined the blackbody
temperature of the CMB to be $2.728 \pm 0.004$ K (Fixsen et. al.
1996)\markcite{fixsen}, and the {\em COBE} DMR has detected spatial anisotropy
of the CMB on $10\arcdeg$ scales of \( \Delta T / T \simeq 1.1 \times 10^{-5} \)
(Bennett et al. 1996)\markcite{benn}.  Ground and balloon based experiments have
also detected anisotropy at smaller scales; see Scott, Silk, \& White
(1995)\markcite{scott} for a review of these results. However, the polarization
of the CMB has received comparatively little experimental attention, despite its
fundamental nature. The anisotropy and polarization depend on the power spectrum
of fluctuations as well as the ionization history of the universe in different
ways. A detection of polarization would complement the detections of anisotropy
by facilitating the reconstruction of the initial spectrum of perturbations as
well as the ionization history of the universe.

The magnitude and spatial distribution of polarization is determined by factors
such as: the source of the CMB anisotropy, the density parameter $\Omega$, the
baryon content of the universe $\Omega_B$, the Hubble constant $H$, and the
ionization history of the universe. CMB polarization is uniquely sensitive to
the ionization history of the universe, which includes the duration of
recombination and the epoch of reionization. The detection of, or a further
constraint on, the polarization of the CMB has the potential to dramatically
enhance our understanding of the pregalactic evolution of the universe.
 
Similar to the CMB anisotropy power spectrum, the polarization power spectrum
contains information on all angular scales.  Large angular scales (larger than
$\simeq 1\arcdeg$) correspond to regions on the last scattering surface which
were larger than the causal horizon. In the absence of reionization, these
scales were affected only by the long wavelength modes of the primordial power
spectrum. This region of the power spectrum was measured by the {\em COBE} DMR,
and establishes the normalization for models of large scale structure
formation. Similarly, measurements of polarization at large angular scales will
normalize the entire polarization power spectrum. Because the anticipated signal
size is small at all angular scales, polarization measurements pose a
significant challenge. While signals from large angular scales may be weaker
than at small scales, the design of a large angular scale measurement is
comparatively simple and compact, with potentially lower susceptibility to
sources of systematic error. A detection, or improved upper limit, at large
angular scales is a natural first step towards probing the polarization power
spectrum on all angular scales.

In this paper we review theoretical arguments which suggest that the ratio of
polarization to anisotropy should be in the range 0.1\% to 10\%, at large
angular scales.  Existing upper limits on polarization are higher than, or
comparable to the measured anisotropy level itself (see Table 1).  Measurements
of anisotropy, by {\em COBE}, and other experiments on the level of \( \Delta T
/ T_{{\rm cmb}} \simeq 1 \times 10^{-5} \) indicate the required level of
sensitivity to polarization must be at least \( \Delta T / T_{{\rm cmb}} \leq 1
\times 10^{-6}\). Thus, to obtain new non-trivial information, either a positive
detection, or an improved upper limit capable of discriminating between
different cosmological scenarios, necessitates extremely precise measurements.

Current detector technology is capable of achieving the required level of
sensitivity. However, in addition to achieving high sensitivity it is essential
to discriminate the polarization from systematic effects, such as
non-cosmological astrophysical sources of polarized radiation.  Space-based
missions, such as MAP\markcite{map} and Planck Surveyor \markcite{cobra} will
produce full-sky anisotropy maps, and are expected to achieve the required
sensitivity level to measure polarization as well. The projected sensitivity
levels will allow for per-pixel detections of anisotropy with signal-to- noise
ratios greater than one.  The polarization maps from these missions, however,
are expected to have signal-to-noise ratios less than one for each beam-sized
pixel, and will be of lower resolution than the anisotropy maps. Fortunately,
polarization observations are also possible from the ground; as we will
demonstrate, polarized atmospheric emission is expected to be negligible.

This paper will concentrate on strategies for a near-term, ground-based
polarization experiment, Polarization Observations of Large Angular Regions
(POLAR), optimized to measure CMB polarization at $7\arcdeg$ scales, for $\sim
36$ pixels.  The design incorporates many techniques developed for previous
anisotropy and polarization experiments, from the ground, balloons, and space.
The primary goal of the paper is to describe the feasibility of measuring large
angular scale polarization, and highlight the conclusions which could be drawn
from such a measurement.  In section \ref{secthy} we review the theory of CMB
polarization which motivates the experimental design. We describe the main
experimental challenges in section \ref{secexpt}, and focus on two which affect
the global design of the instrument: discrimination of CMB polarization from
polarized foreground sources (in sections \ref{secfor} and \ref{secsub}), and an
observing strategy designed to minimize the time required to detect a
cosmological signal (section \ref{secobs}). Finally, we estimate the
polarization signal we expect in several different cosmological scenarios, and
speculate on the conclusions which could be drawn from such a detection.

\section{CMB Polarization: Theory}
\label{secthy}
Anisotropy of the CMB is generated by metric perturbations of the universe.
There are two primary types of perturbation that generate anisotropy of the CMB:
scalar contributions, generated by matter and radiation density inhomogenities,
and tensor contributions, associated with gravitational waves.  Both types of
perturbation give rise to temperature fluctuations in the CMB via the
Sachs-Wolfe effect (Sachs \& Wolfe 1967). \markcite{sachs}

Thomson scattering of anisotropic radiation by free electrons inevitably
generates polarization (Chandrasekhar 1960)\markcite{chandra}. Scattering by a
single electron produces polarized radiation with an intensity approximately
10\% of the anisotropy quadrupole amplitude when averaged over all directions of
photon incidence and scattering. In the case of CMB polarization, the exact
polarization level, as well as the angular scale of the distribution of
polarization on the sky depend on the optical depth along the observer's line of
sight, and on the particular sources of metric perturbation (Rees 1968; Basko \&
Polnarev 1980; Negroponte \& Silk 1980; Tolman
1985)\markcite{rees}\markcite{basko}\markcite{neg}\markcite{tol}.  For a recent
review see: Hu (1996), Kosowsky (1996)\markcite{hu}\markcite{kos}.

According to the standard model of the evolution of the pre-galactic medium
after recombination, the previously ionized plasma formed neutral hydrogen which
was transparent to the CMB. However, the universe may have undergone a secondary
ionization of the recombined hydrogen.  Gunn \& Peterson (1965) \markcite{gunn}
formulate a measurement of the ionization fraction of the intergalactic medium
using the lack of a Lyman-$\alpha$ trough in the observed spectra of distant
quasars. Recent results show that the majority of intergalactic hydrogen is
highly ionized to a redshift of at least $z \sim 5$, indicating that the
universe must have reionized at an earlier epoch (Peebles 1993). Several models
predict reionization occurred in the redshift range approximately $30 < z_{{\rm
r i}} < 70 $ (Ozernoi \& Chernomordik 1976; Gooding et al. 1991; Durrer 1993;
Tegmark \& Silk 1993; Nasel'skii \& Polnarev 1987).
\markcite{oz}\markcite{nas}\markcite{oz}\markcite{good}\markcite{dur}\markcite{tegmark}

In contrast to the standard model of recombination, non-standard models invoke
additional non- equilibrium sources of ionization. These models predict a
prolonged, or even non-existent, recombination and/or subsequent ionization of
the recombined plasma. Since polarization is generated by scattering of photons
on free electrons, its magnitude and spatial distribution could be used to
discriminate between non-standard models and the standard model (Bond \&
Efstathiou 1984,1987; Basko \& Polnarev 1980; Nasel'skii \& Polnarev 1987; Ng \&
Ng 1996; Zaldarriaga \& Harrari 1995; Crittenden, Davis, \& Steinhardt 1993;
Frewin, Polnarev, \& Coles 1994).\markcite{bond}\markcite{basko}\markcite{nasel}
\markcite{ng3}\markcite{zal}\markcite{crit2}\markcite{frew} An early
reionization effectively introduces an additional `last' scattering
surface. This has two effects, both of which, in principle, can enhance the
magnitude of the polarization on large angular scales. Primarily, the additional
scattering of photons during reionization can create new, or amplify existing
polarized radiation via the Thomson mechanism discussed above. Additionally, the
second `last' scattering surface occurs at a much lower redshift, implying that
the causal horizon on this rescattering surface is larger, and thus, will
subtend a larger angle on the sky today.

In general, large-scale polarization is enhanced in models which predict early
reionization. As we will demonstrate, for reasonable non-standard models, the
amplitude of polarization on $10\arcdeg$ angular scales is on the level of
$10\%$ of the anisotropy, while for the standard model of recombination the
corresponding polarization level does not exceed $1\%$. It is worth mentioning
that all of these models predict approximately the same level of anisotropy at
$10\arcdeg$ scales, and hence all of them are compatible with the results of the
{\it COBE} DMR experiment.

In the remainder of this section we will illustrate the important theoretical
 features of the polarization of the CMB.  We will first describe an analytic
 treatment which predicts the level of polarization for both standard and
 non-standard reionization h istories. In section 2.3 we will describe a
 numerical simulation of the effect of a non-standard reionization history on
 the polarization of the CMB. We will find that the more qualitative analytic
 results agree quite well with the quantitative results of our numerical
 simulations.

\subsection{Polarization produced by cosmological perturbations}

Here we develop the mathematical formalism which will allow us to describe the
polarization of the CMB in a consistent fashion. With these tools, we will
subsequently determine the polarization signal we expect to observe, using two
different techniques.  The first method is an analytic approach which will
provide a physical framework for understanding the polarization of the CMB. The
second approach is more quantitative, and will allow us to obtain numerical
estimates of the polarization signal. In order to describe the polarization of
the CMB, we will first introduce a parameterization which describes the
polarization state of arbitrary radiation fields. We then apply this formalism
to the polarization state of the cosmological signal which we are seeking to
detect.

Consider a polarized electromagnetic wave with angular frequency, $\omega$:
${\bf \vec{E}} = E_{y0}\sin (\omega t - \delta _y){\bf \hat{y}} + E_{x0}\sin
(\omega t - \delta _x){\bf \hat{x}}$. The polarization state of electromagnetic
radiation can be characterized by the Stokes parameters: $I, Q, U,$ and $V$.  $I
= I_y + I_x $, with $I_y = \lan E_{y0}^2\ran$ and $ I_x = \lan
E_{x0}^2\ran$. $I$ is the total intensity of the radiation, and is always
positive. The parameters $Q = I_y - I_x$ and $U = 2E_{y0}E_{x0}\cos (\delta _y -
\delta_x)$ quantify the linear polarization of the wave, and $V$ quantifies the
degree of circular polarization (when $V=0$, the radiation is linearly polarized
or unpolarized). The level of polarization is defined as $ \Pi = \frac{\sqrt{Q^2
+ U^2 + V^2}}{I}$, and the polarized intensity is $I_{{\rm pol}} \equiv \Pi
\times I$.

An alternate representation for the Stokes parameters will be of use in the
following sections. We introduce a symbolic vector for the distribution function
of occupation numbers of polarized radiation: ${\bf \hat{n}} = \frac{c^2}{h
\nu^3}{\bf \hat{I}}$, where ${\bf \hat{I}}$ is the symbolic vector introduced in
Chandrasekhar (1960) and related to the Stokes parameters in the following way:

\[
{\bf \hat{I}} = \left( \begin{array}{c} I_{x} \\ I_{y} \\ U \\ V \end{array} \right)
.\]

Since Thomson scattering cannot produce circular polarization, $V=0$, we will
consider the 3-vector: ${\bf \hat{I}} = \left( \begin{array}{c} I_x \\ I_y \\ U
\end{array} \right).$ An unpolarized distribution in zero-th order approximation
is given by ${\bf \hat{n}_o} = n_o \left( \begin{array}{c} 1 \\ 1 \\ 0
\end{array} \right)$.

As shown in (Basko \& Polnarev 1980)\markcite{basko} and further discussed in
(Polnarev 1985; Zaldarriaga \& Harrari 1995)\markcite{pol85}\markcite{zal},
polarized radiation in the presence of cosmological perturbations can be
represented as

\be
{\bf \hat{n}} = n_o\left[ \left( \begin{array}{c} 1 \\ 1 \\ 0 \end{array}
\right) + {\bf\hat{n}_1} \right],
\ee

where ${\bf \hat{n}_1} = {\bf \hat{n}_A}+ {\bf \hat{n}_{\Pi}}$ is the correction
to the uniform, isotropic, and unpolarized radiation described by ${\bf
\hat{n}_o}.$ The Planck spectrum, ${\bf \hat{n}_o}$, depends only on frequency,
and ${\bf \hat{n}_A}+ {\bf \hat{n}_{\Pi}}$ are the anisotropic and polarized
components, respectively, which are functions of the conformal time, $\eta$,
comoving spatial coordinates, $x^{\alpha}$, photon frequency, $\nu$, and photon
propagation direction specified by the unit vector $\hat{e}(\theta,\phi)$ with
polar angle, $\theta$, and azimuthal angle, $\phi$, given in an arbitrarily
oriented spherical coordinate system.

The Equation of Radiative Transfer in terms of ${\bf \hat{n}}(\eta,x^{\alpha},\nu,\mu,\phi)$, where $\mu 
= \cos \theta$, is:

\be
\frac{ \partial {\bf \hat{n}}}{\partial \eta} + {\bf e^{\alpha}} \cdot \frac {\partial {\bf \hat{n}}}{\partial 
x^{\alpha}} = -\frac{\partial {\bf \hat{n}}}{\partial \nu} \frac{\partial {\nu}}{\partial \eta} - q( {\bf 
\hat{n}} - {\bf \hat{J}}) \label{eqert}
\ee
and
\be
{\bf  \hat{J}} = \frac{1}{4\pi}\int^{+1}_{-1}\int^{2\pi}_{0}  {\bf \hat{\hat{P}}} 
(\mu,\phi,\mu^{\prime},\phi^{\prime}) {\bf \hat{n}} 
(\eta,x^{\alpha},\nu,\mu\arcmin,\phi\arcmin)d\mu^{\prime} d\phi^{\prime},
\ee

where $q = \sigma_T N_e a$, and the Einstein summation convention is implied. In
these expressions, $a$ is the cosmological scale factor, ${\bf \hat{\hat{P}}}$
is the scattering matrix described by Chandrasekhar (1960), $\sigma _T$ is the
Thomson cross section, and $N_e$ is the comoving number density of free
electrons. In general, the effects of a particular choice of metric perturbation
are manifest in the first term on the right hand side of (\ref{eqert}):

\[
\frac{\partial {\nu}}{\partial \eta}= \frac{1}{2} \frac{\partial h_{\alpha\beta}}{\partial \eta}  
e^{\alpha}e^{\beta}\nu
\] (Sachs \& Wolfe 1967)\markcite{sachs}. After retaining terms up to first order in 
metric perturbations, $h_{\alpha\beta}$, and since $ \frac{\partial
{\nu}}{\partial \eta}$ is of the first order, we can replace $\frac{\partial
{\bf \hat{n}}}{\partial \nu}$ by $ \frac{\partial {\bf \hat{n}_o}}{\partial
\nu_o}$ in the source term ($\nu_o$ is the unperturbed frequency). This implies
that the factor \[ \gamma = \frac{\nu_o}{n_o} \frac{dn_o}{d\nu_o}=\frac{d\ln
n_o}{d\ln\nu_o}\]gives a universal frequency dependence for anisotropy and polarization effects,
independent of the type of metric perturbations (Basko \& Polnarev
1980)\markcite{basko}.

The angular dependence of ${\bf \hat{\hat{P}}}$ is such that

\be
\frac{1}{4\pi}\int^{+1}_{-1}\int^{2\pi}_{0}{\bf \hat{\hat{P}}} 
(\mu,\phi,\mu^{\prime},\phi^{\prime}){\bf \hat{n}_o}d\mu^{\prime} d\phi^{\prime}=\hat{0},
\ee

(where $\hat{0}$ is the symbolic zero-vector), so we conclude in the zero-th
order approximation, ${\bf \hat{J}} = \hat{0}$. For the first order
approximation, in the following, we will understand by ${\bf \hat{J}}$ actually
${\bf \hat{J}_1}$, in which ${\bf \hat{n}}$ is replaced by ${\bf \hat{n}_1}$.

After linearization and spatial Fourier transformation, the equation of transfer
takes the following form (with $\nu_o$ replaced by $\nu$):

\be
\frac{\partial {\bf \hat{n}}_{1\vec{k}}}{\partial \eta} + ik\mu {\bf \hat{n}}_{1\vec{k}} = \gamma 
H_{\vec{k}} - q ( {\bf \hat{n}}_{1\vec{k}} - {\bf \hat{J}_{\vec{k}}}).
\ee

Here, $H_{\vec{k}} =-\frac{1}{2}\dot{h}_{\alpha\beta\vec{k}}
e^{\alpha}e^{\beta}$, and ``$\dot{\hspace{0.1in}}"\equiv \frac{d}{d\eta}$. We
have specified spherical coordinates in such a way that $\mu = \cos\theta$,
where $\theta$ is the angle between a vector along the line of sight, ${\bf
\hat{e}}$, and the wave vector $\vec{k}$ and $\phi$ is the azimuthal angle of
the vector ${\bf \hat{e}}$, in the plane perpendicular to the vector $\vec{k}$.

For a given $\vec{k}$, $h_{\alpha\beta\vec{k}}$ can be represented as a
superposition of scalar waves (below we will use subscript ``S") and tensor
gravitational waves (subscript ``T" ). Taking into account the tensorial
structure of the waves, and restricting our consideration to perturbations with
wavelengths longer than the cosmological horizon at the moment of equipartion
(i.e. at the moment when the energy density of matter equals that of radiation,
see for example Harrari \& Zaldarriaga (1993)\markcite{har1}), we can write

\be
H_{\vec{k}} = \frac{1}{15}\eta k^2 \mu^2 \kappa_S (k) -
\frac{3}{2k^3}(1-\mu^2)\cos 2\phi
\frac{d}{d\eta}\left[\frac{1}{\eta}\frac{d}{d\eta}\left( \frac{\sin
k\eta}{\eta}\right)\right] \kappa_T(k) \ee
 Here, the $\sqrt{\vert
\kappa_{S,T}(k)\vert ^2}$ are the amplitudes of the corresponding metric
perturbations at the moment when their wavelengths are equal to the cosmological
horizon.

For $k\eta \ll 1$ we have:

\be
H_{\vec{k}}  \simeq  \frac{1}{15}\eta k^2 \mu^2 \kappa_S (k) - \frac{3}{2}(1-\mu^2)\cos 2\phi  
\kappa_T(k)
\label{eqfu}
\ee

while for $k\eta \gg 1$, 

\be
H_{\vec{k}}  = \frac{1}{15}\eta k^2 \mu^2 \kappa_S (k) + \frac{3}{k \eta^2}(1-\mu^2)\cos 2\phi \cos 
k\eta \kappa_T(k)
\ee
 
For a plane wave perturbation with wavevector $\vec{k}$, the anisotropy and
polarization can be described as (Basko \& Polnarev 1980)\markcite{basko}:

\be
{\bf \hat{n}_A} = \alpha_S (\mu^2 - \frac{1}{3}) \left( \begin{array}{c}1\\1\\0 \end{array} \right) + 
\frac{\alpha_T}{2}(1-\mu^2)\left( \begin{array}{c}1\\1\\0 \end{array} \right)\cos 2\phi \label{eqani}
\ee

\be
 {\bf \hat{n}_{\Pi}} = \beta_S(1 - \mu^2 ) \left( \begin{array}{c} 1 \\ -1 \\ 0 \end{array} \right) + \beta_T 
\left( \begin{array}{c}(1+\mu^2)\cos 2\phi \\ -(1+\mu^2)\cos 2\phi \\ 4\mu \sin 2\phi  \end{array} \right) 
\label{eqpola}
\ee

Substituting (\ref{eqani}) and (\ref{eqpola}) into the integro-differential
Equation of Radiative Transfer, (\ref{eqert}), we obtain the following system of
coupled ordinary differential equations for $\alpha_{S,T}$ and $\beta_{S,T}$:

\be
\dot{\beta}_{S,T} + \frac{3}{10} q \beta_{S,T} = -\frac{1}{10}q \xi_{S,T}
\ee

\be
\dot{\xi}_{S,T} + q \xi_{S,T} = F_{S,T},
\ee

where $\xi_{S,T} = \alpha_{S,T} + \beta_{S,T}$, and $F_{S,T}$ is the appropriate
source function for scalar or tensor perturbations. This system of coupled
equations illustrates the intimate relation between anisotropy and the
generation of polarization.  Integrating this system of equations we obtain the
following general solution for $\beta_{S,T}$:

\be
\beta_{S,T} = \frac{1}{7}\int_0^{\eta} F_{S,T} \left[ e^{-\tau} - e^{-\frac{3}{10}\tau} \right] 
d\eta^{\prime}
\label{eqbs}
\ee 

where \(\tau(\eta,\eta')= \int^{\eta}_{\eta\arcmin}q(x^{\alpha})dx^{\alpha}\) is
the optical depth with respect to Thomson scattering.

For wavelengths large in comparison with the cosmological horizon at the moment of decoupling, 
$\eta_D$, ($k\eta_D \ll 1$), the source function, at this moment can be approximated by:

\be
F_{S,T} = \frac{\gamma}{15}\eta k^2 \left\{ \begin{array}{c} \kappa_S (\vec{k})  \\ -\frac{2}{3} 
\kappa_{T} (\vec{k})  \end{array} \right.
\ee

It can be shown that the source functions are rather insensitive to the exact
functional form of the variation of the optical depth with respect to
time. (Basko \& Polnarev 1980; Nasel'skii \& Polnarev
1987)\markcite{basko}\markcite{nasel} These functions are primarily
characterized by the epoch and duration of decoupling.  Following (Zaldarriaga
\& Harrari 1995)\markcite{zal}, we adopt the following approximation for the
time variation of the optical depth:
\[d\tau = - \frac{d\eta}{\Delta\eta_D}\tau\] (see  also (Basko \& Polnarev 1980; Nasel'skii \& Polnarev 
1987) for a more detailed discussion.). Here $\Delta\eta_D$ is the characteristic time scale of the duration 
of decoupling. Approximating the source functions under the integral (\ref{eqbs}), by their values at the 
moment of decoupling $\eta_D$, which gives the main contribution to polarization, we have

\be
\beta_{S,T} \simeq \frac{1}{7}(F_{S,T})\vert_D \Delta\eta_D\int_0^\infty \left[ e^{-\tau} - e^{-
\frac{3}{10}\tau} \right] \frac{d\tau}{\tau}
\label{eqbe}
\ee
The integral in (\ref{eqbe}) can be evaluated in the following way:
\be
\beta_{S,T} \simeq \frac{1}{7}(F_{S,T})\vert_D \Delta\eta_D \lim_{\epsilon\rightarrow 0} \left( 
\int_{\epsilon}^\infty  e^{-\tau}\frac{d\tau}{\tau} - \int_{-\frac{3}{10}\epsilon}^\infty e^{-
\tau\arcmin}\frac{d\tau\arcmin}{\tau\arcmin}\right)
\ee

\[= \frac{1}{7}\ln\frac{10}{3}(F_{S,T})\vert_D \Delta\eta_D\]

Hence,

\be
{\bf \hat{n}} = -\frac{1}{105}\ln\frac{10}{3}\eta_D \Delta\eta_D k^2\gamma \left\{\frac{1}{2}\kappa_S 
(k)(1-\mu^2) \left( \begin{array}{c}1\\-1\\0 \end{array} \right) - \frac{3}{4}\kappa_T(k) \left( 
\begin{array}{c} (1+\mu^2) \cos2\phi \\ -(1+\mu^2)\cos 2\phi \\ 4\mu\sin 2\phi \end{array} 
\right)\right\}\label{eqfinalpol}
\ee  

Comparing equation (\ref{eqfinalpol}) with equation (\ref{eqpola}), we find that
the polarization generated by a single perturbation mode with wavevector
$\vec{k}$ is given by:

\be
\Pi_{\vec{k}} = - \frac{2}{105} \ln \frac{10}{3} (\eta_D k) (\Delta_D k) \gamma \left\{\kappa_S(k) (1-
\mu^2) - \frac{3}{4}\kappa_T(k)\left[(1+\mu^2) \cos 2\phi + 2\mu \sin 2\phi \right] \right\}.
\label{eqpik}
\ee

Now we can calculate the root mean square (rms) polarization measured by an
antenna with an effective averaging angle $\Theta_A$. The main contribution to
the rms polarization, $\Pi(\Theta_A)$ , is contributed by modes with $k <
k_{{\rm max}}(\Theta_A) \approx \frac{2\pi}{\Theta_A} = \frac
{360\arcdeg}{\Theta_A\arcdeg}$:

\be
 \Pi(\Theta_A) = \sqrt{\lan\Pi^2 \ran _{k > \frac{2\pi}{\Theta_A}}} 
\ee

\be
= \frac{2}{105} \ln\frac{10}{3}\eta_D\Delta\eta_D \gamma \sqrt{Q_S B_S+ Q_T B_T },
\ee

where:

\begin{eqnarray} B_S &=& \int_{-1}^{1}(1-\mu^2)^2d\mu = \frac{16}{15} \\ B_T &=& 
\frac{9}{8\pi}\left[\int^1_{-1} (1+\mu^2)^2\int_0^{2\pi}\cos^2 2\phi d\phi + \int^1_{-
1}4\mu^2d\mu\int_0^{2\pi}\sin^2 2\phi \right] =  \frac{36}{5}.
\end{eqnarray} 

Here $Q_{S,T} = \int^{k_{\max}(\Theta_A)}_0 k^4 \vert \kappa_{S,T}(k)\vert ^2
\frac{dk}{k}$, with $\vert \kappa_{S,T}(k)\vert^2 =
\kappa_{0_{S,T}}k^{n_{S,T}}$, and $\sqrt{\vert\kappa_{0_{S,T}}\vert^2}$ are the
amplitudes of perturbations with wavelengths equal to the cosmological horizon
at the present moment ($n=0$ corresponds to a scale invariant Harrison-Zeldovich
spectrum). These amplitudes are normalized to the {\it COBE} DMR anisotropy
quadrupole detection which is approximately equal to $2 \times 10^{-5}$.
Assuming that $n_S = n_T = n$, we obtain \be \Pi(\Theta_A) =
\frac{8}{105\sqrt{15}}\ln\frac{10}{3}\eta_D\Delta\eta_D \frac{1}{\sqrt{4+n}}
\left( \frac{360\arcdeg}{\Theta_{A}\arcdeg}\right)^{2+\frac{n}{2}}\gamma
\sqrt{\kappa^{2}_{o_{S}} + \frac{27}{4}\kappa^2_{o_T}}
\label{eqfu2}
\ee

Taking into account the relationship between redshift and conformal time, $z
\sim \frac{1}{\eta^2}$, we have that $\frac{\Delta z_D}{z_D} \sim
\frac{2\Delta\eta_D}{\eta_D}$, hence
\[ \eta_D\Delta\eta_D \simeq \eta_D^2 \frac{\Delta\eta_D}{\eta_D} = \frac{1}{2} \frac{\Delta 
z_D}{z_D^2} = \frac{1}{2} \frac{\Delta
z_D}{z_D}\frac{1}{z_{SD}}\left(\frac{z_{SD}}{z_D}\right)
\]
where $z_{SD}$ is the redshift of decoupling predicted by the standard model of
recombination. Finally:
\be
\Pi(\Theta_A) = 4 \times 10^{-7} \frac{\Delta z_D}{z_D} \left( \frac{z_{SD}}{z_D} \right) \left( 
\frac{7\arcdeg}{\Theta_A} \right) ^{2+\frac{n}{2}} \gamma\aleph_{n,g} \label{eqpol}
\ee
where 
\be
\aleph_{n,g} = \frac{2\times 10^{-2} \ln\frac{10}{3}}{105\sqrt{15}} \left( \frac{7\arcdeg}{360\arcdeg} 
\right) ^2 \frac{\sqrt{\kappa^2_{o_T} + \kappa^2_{o_S}}}{2\times 10^{-5}} \frac{10^3}{z_{SD}} \left( 
1+\frac{n}{4} \right) ^{-1/2} \left( \frac{360\arcdeg}{7\arcdeg} \right) ^{\frac{n}{2}} \sqrt{\frac{1 + 
\frac{27}{8} g^2}{1+g^2}}
\label{eqfu3}
\ee
and \[g = \frac{\kappa_{o_T}}{\kappa_{o_{S}}}\] is the ratio of the tensor
perturbation amplitudes to the scalar amplitudes.

The factor $\aleph_{n,g}$ incorporates the perturbation amplitudes, normalized to the anisotropy 
quadrupole measured by the {\em COBE} DMR. It contains all information about the type of metric 
perturbation, allowing us to isolate factors which depend upon the nature of the perturbations, and those 
which do not. For $n=0$, and $g=0$ (i.e., no tensor perturbations), $\aleph_{o,g} \simeq 1$. When 
$g=\infty$ (i.e., no scalar perturbations), $\aleph_{o,g} \simeq 1.84$. Finally, when $g=1$ (i.e., equal 
tensor and scalar contributions), we find  $\aleph_{o,g} \simeq 1.47$. From this we observe that
 $\aleph_{n,g}$ is rather insensitive to the ratio of tensor to scalar amplitudes, g.

We now emphasize the angular regions to which the preceding discussion is relevant. Equations 
(\ref{eqfu2}) - (\ref{eqfu3}) ( which are based on asymptotic formula (\ref{eqfu}), and the 
approximations used in (\ref{eqbe})), are valid for modes which satisfy: $k\Delta\eta_D < 1$. In terms of 
angle on the sky, \[\frac{360\arcdeg}{\Theta_A\arcdeg}\frac{\Delta\eta_D}{\eta_D}\eta_D < 1.\]

We can apply equations  (\ref{eqfu2}) - (\ref{eqfu3}) to an observation which has an angular resolution 
$\Theta_A$, as long as: 

\[\Theta_A > \Theta_{A_{{\rm min}}} = 360\arcdeg \frac{1}{2} \frac{\Delta z_D}{z_D^{3/2}} = 
\frac{180}{z_{SD}^{1/2}} \left( \frac{\Delta z_D}{z_D} \right) \left( \frac{z_{SD}}{z_D} \right) ^{1/2} 
\simeq 6\arcdeg \frac{\Delta z_D}{z_D} \left( \frac{z_{SD}}{z_D} \right )^{1/2} \left( \frac{\Delta 
z_{SD}}{10^3} \right) ^{-1/2}.\]

As an example, the standard model of recombination predicts $\frac{\Delta z_D}{z_D} \simeq 0.1$, 
which implies that $\Theta_{A_{min}} \simeq 0.6\arcdeg$. For pure scalar perturbations ($n=0$), the 
expected level of polarization at this angular scale is: $\Pi(0.6\arcdeg) \simeq 6\times 10^{-6}$ . For an 
observation with $ \Theta_{A} \simeq 6\arcdeg $, the polarization is $\Pi(6\arcdeg) \simeq 5 \times 10^{-
8}$. The observed polarization is suppressed by a factor of $\sim 100$ with this lower resolution beam. 

Consider another example, a non-standard model for which $\frac{\Delta z_D}{z_D} \simeq 1$, and  
$z_D \simeq z_{SD}$, the angular scale is:  $ \Theta_{A_{min}} \simeq 6\arcdeg $. The polarization 
predicted in this scenario is: $\Pi(6\arcdeg) \simeq 5 \times 10^{-7}$. Finally, for  $ \Theta_{A}< 
\Theta_{A_{min}}$, the polarization is suppressed, and its dependence on $\Theta_A$ is determined by 
the details of the ionization history (Zaldarriaga 1997; Polnarev 1985; Nasel'skii \& Polnarev 1987; Bond 
'\& Efstathiou 1984,1987)\markcite{zalsolo}\markcite{pol85}\markcite{nasel}\markcite{bond}.

To summarize, for a given $\Theta_A$, the polarization level is proportional to $\frac{\Delta 
z_D}{z_D}$, (see equation (\ref{eqpol})) and is smallest for the standard model of recombination. 
Alternatively, this analytic approximation applies to smaller angles in the standard model, as opposed to 
the larger angles predicted by non-standard models (see Nasel'skii \& Polnarev (1987)\markcite{nasel} for 
a more detailed discussion. Figure \ref{figthy1} schematically illustrates the angular dependence of 
polarization in standard and non-standard models. 

\placefigure{figthy1}

\subsection{Polarization Power Spectrum}
The analytic treatment above describes the essential physics responsible for the generation of CMB 
polarization. We have discussed the aspects of non-standard recombination which are relevant to the large 
scale polarization of the CMB. In order to estimate the observable polarization signature, we now detail a 
more quantitative approach based on the polarization power spectrum. This approach also allows us to 
discuss the effect of an early reionization on the observed polarization.

For quantitative estimates, the polarization and anisotropy source terms which appear in the equation of 
transfer can be decomposed into Legendre series. The individual modes are then evolved to the present 
where the spatial structure  of the CMB can be computed (see Bond \& Efstathiou 1984,1987; Ng \& Ng 
1995, Zaldarriaga \& Harrari 1995; Frewin, Polnarev, \& Coles 1994, for 
example)\markcite{bond}\markcite{zal}\markcite{frew}. Because the CMB is an imprint of the epoch of 
linear evolution of perturbations, the individual modes evolve more or less independently. This treatment 
lends itself particularly well to numerical analysis  (Seljak \& Zaldarriaga 1996a)\markcite{seljak}. The 
relevant results of such analysis to the present discussion are the anisotropy and polarization power 
spectra. Here we connect the results of these numerical procedures with the analytic treatment presented in 
the previous subsection.

The temperature of the CMB, being a scalar valued function, can be expanded in a
spherical harmonic series on the sky, at a particular point on the sky,  ${\bf \hat{x}}$:
\be
T(({\bf \hat{x}}) = \sum_{\ell,m} a_{T,\ell m}Y_{\ell m}({\bf \hat{x}})
\label{eqtsolo}
\ee
where the $Y_{\ell m}({\bf \hat{x}})$ are the spherical harmonics at ${\bf \hat{x}}$. The temperature 
two-point correlation function is given by:
\be
C_{T,\ell} = \frac{1}{2\ell +1} \sum_{m} \lan a_{T,\ell m}^{*}a_{T,\ell m} \ran \label{eqtcorr}
\ee

The variance of the $a_{T,\ell m}$ is given by the $C_{T,\ell}$, since Var$[a_{T,\ell m}] = \lan\vert 
a_{T,\ell m}\vert ^2\ran - \lan  \vert a_{T,\ell m}\vert \ran ^2 = \lan \vert a_{T,\ell m}\vert ^2\ran \equiv 
C_{T,\ell}$
if the  $a_{T,\ell m}$ are gaussian distributed with zero mean, and $\lan\ldots\ran$ denotes a whole-sky 
average followed by an average over all observational positions.

The polarization of the CMB is a tensor-valued function, with a symmetry group different from that of the 
anisotropy. As shown in (Zaldarriaga \& Seljak 1997), complex, linear combinations of the Stokes 
parameters transform under rotation about the line of sight by an angle $\psi$ as:
\be
(Q \pm iU)\arcmin({\bf \hat{x}}) = \exp^{\mp 2i\psi}(Q \pm iU)({\bf \hat{x}})
\ee
The analogous expressions to equation \ref{eqtsolo} for the Stokes parameters are:

\begin{eqnarray}
(Q + iU)({\bf \hat{x}}) &=& \sum_{\ell,m} a_{2,\ell m}^{\Pi} \hbox{}_{2}Y_{\ell m}({\bf \hat{x}})\\
(Q - iU)({\bf \hat{x}}) &=& \sum_{\ell,m} a_{-2,\ell m}^{\Pi} \hbox{}_{-2}Y_{\ell m}({\bf \hat{x}})
\end{eqnarray}

where the spin-weighted spherical harmonics, $_{-2}Y_{\ell m}$, are a complete and orthonormal set of 
basis functions on the sphere (for an equivalent technique, which uses second rank tensors on the sphere to 
describe the Stokes parameters, see (Kamionkowski et al. 1996)). Taking complex, linear combinations of 
the expansion coefficients, $a_{\pm2,\ell m}^{\Pi}$, one defines:
\begin{eqnarray}
 a_{E,\ell m} &\equiv&  -(a_{2,\ell m}^{\Pi} +  a_{-2,\ell m}^{\Pi})/2\\
 a_{B,\ell m} &\equiv&  i(a_{2,\ell m}^{\Pi} -  a_{-2,\ell m}^{\Pi})/2.
\end{eqnarray}
From these we construct two independent correlation functions which characterize the polarization:

\begin{eqnarray}
C^{\Pi}_{E,\ell} = \frac{1}{2\ell +1} \sum_{m} \lan a_{E,\ell m}^{*}a_{E,\ell m}\ran \\
C^{\Pi}_{B,\ell} = \frac{1}{2\ell +1} \sum_{m} \lan a_{B,\ell m}^{*}a_{B,\ell m}\ran 
\end{eqnarray}

The $C^{\Pi}_{E,\ell}$ and $C^{\Pi}_{B,\ell}$ have different physical origins and have interesting 
properties under symmetry transformations, such as parity inversion. We refer to (Seljak \& Zaldarriaga 
1996b) for further discussion of these fascinating spectra. 

We now define $C_{\ell}^{\Pi} \equiv C^{\Pi}_{E,\ell} + C^{\Pi}_{B,\ell}$  and form:
\be
\lan Q({\bf \vec{x}_1})Q({\bf \vec{x}_2}) + U({\bf \vec{x}_1}) U({\bf \vec{x}_2}) \ran = 
\frac{1}{4\pi}\sum_{\ell=2}^{\infty}(2\ell + 1)C_{\ell}^{\Pi} P_{\ell}(\cos\theta)
\ee
where ${\bf \vec{x}_1}$ and ${\bf \vec{x}_2}$ are vectors toward two different locations on the sky 
separated by an angle $\theta$. When $\theta = 0$ we have for the polarization autocorrelation:

\[ I^2_{\rm pol} = \lan Q^2 + U^2 \ran = \frac{1}{4\pi}\sum_{\ell=2}^{\infty}(2\ell + 1) C_{\ell}^{\Pi}
\]

We can now connect the results of this power spectrum calculation with the analytic treatment presented 
above. For a measurement of the polarization at an angular scale $\Theta_A$, we have for the observable 
level of polarization:
\be
\Pi (\Theta_A) \equiv \frac{I_{{\rm pol}}}{I} =\frac{1}{I}
 \sqrt{\frac{1}{4\pi}\sum_{\ell=2}^{\infty}(2\ell + 1)W_{\ell}^{\Theta_A} \times 
C_{\ell}^{\Pi}},\label{polopol}
\ee
where $I \simeq 2.7$K, and the window function, $W_\ell^{\Theta_A}$, is a factor
which weights the contribution of the $\ell^{{\rm th}}$ moment to the power
spectrum. It quantifies the angular sensitivity of a given experiment (see
section \ref{secobs} for further detail). For a given theoretical model
(i.e. $C_{\ell}$), equation \ref{polopol} can be compared to the analytic
expression given in equation \ref{eqpol}.

The polarization power spectrum, $C_{\ell}^{\Pi}$, is highly dependent on the cosmological details of the 
model used to generate the anisotropy spectrum, $C^{\alpha}_{\ell}$. For this reason, the ratio of 
polarization to anisotropy is often calculated in order to predict observable levels of polarization for 
particular observations  (Ng \& Ng 1995; Crittenden, Davis, \& Steinhardt 1993). In practice, the 
polarization spectra are normalized to the appropriate anisotropy spectra, which are in turn normalized to 
the low $\ell$ values of $C_{\ell}^{\alpha}$ measured by the {\it COBE} DMR.  The power spectra are 
generated numerically by  CMBFAST (Seljak \& Zaldarriaga 1996a)\markcite{seljak}, which permits 
simultaneous calculation of anisotropy and polarization, as well as consistent normalization.

\subsection{The Effect of Reionization on the Polarization Power Spectrum}

As mentioned earlier, non-standard models of the ionization history are characterized by non-
instantaneous decoupling and/or non-zero optical depth along CMB photon trajectories. We have 
discussed the effect a of non-instantaneous recombination using the analytic method treated above. We 
now wish to examine the effect of reionization on the details of the polarization of the CMB. This 
investigation lends itself particularly well to the numerical evaluation of the polarization power spectrum, 
calculated using numerical routines such as CMBFAST. 

In general, models of reionization often rely on structures such as an early generation of stars (Population 
III), or energetic proto-galaxies to provide either ionizing radiation or collisional heating mechanisms. 
Thus, for every model of reionization there corresponds a structure formation scenario, as well as a 
commensurate set of cosmological parameters to be confronted with observational evidence. We will not 
speculate here on the plausibility of specific models of reionization. Discussion of mechanisms for early 
reionization can be found in (Ozernoi \& Chenomordik 1976;  Gooding  et al. 1991; Durrer 1993; Tegmark 
\& Silk 1993; Nasel'skii \& Polnarev 1987). As noted above, the Gunn-Peterson test provides definitive 
evidence for an ionized intergalactic medium out to a redshift of at least $z=5$. In fact, the upper limit on 
the redshift of reionization is set only by the paucity of observed quasars beyond $z=5$ and, in principle, 
could be much higher than this. The {\em COBE} FIRAS limit on the Compton-$y$ parameter  $ y = \int 
d\tau k_b(T_e - T_{{\rm cmb}})/m_e c^2  \leq 2.5 \times 10^{-5} $  (Fixsen et al. 1996), severely restricts 
the energy input allowed in models of reionization, but does not tightly constrain the epoch of reionization 
or the ionized fraction of the intergalactic medium. The limit is compatible with many early reionization 
scenarios.

The effect of reionization can be parameterized in two equivalent forms. One method is specified by the 
optical depth, $\tau_{{\rm ri}}$, for photons due to Thomson scattering along a line of sight to the last 
scattering surface. The second method specifies the redshift of reionization, $z_{{\rm ri}}$, and the 
fractional ionization $x$ (electron-to-proton ratio). The two parameterizations are related as follows 
(Peebles 1993)\markcite{peebles}:

\be
\tau_{{\rm ri}}= 0.0015\left( \frac{x}{1}\right) \frac{\Omega_B}{0.05} \left( \frac{\Omega}{1}\right)^{-
1/2} \left( \frac{h}{0.65} \right) (1+z_{{\rm ri}})^{3/2}
\label{eqtauvzri}
\ee

where $h$ is the Hubble parameter, $\Omega$ is the total density parameter of the universe, and 
$\Omega_B$ is the density parameter of baryonic matter. Equation (\ref{eqtauvzri}) shows the effect of 
curvature of the universe on the optical depth. For reionization occurring at the same redshift and 
ionization fraction, in an open universe ($\Omega < 1$), the optical depth will be greater than in a flat or 
closed universe. We also note that the physical size of regions which are in causal contact (Hubble radius) 
at the epoch of reionization, $t_{{\rm ri}}$, is of order $\sim c t_{{\rm ri}}$. We expect that regions of 
this size will produce coherent polarization of the CMB, and affect the observed polarization power 
spectrum at angular scales which  correspond to the angular scale subtended by the horizon size at the 
epoch of reionization. This argument is  similar to those which predict a coherence scale in the CMB 
anisotropy power spectrum. For example, the acoustic peaks in the CMB anisotropy power 
spectrum arise from causal mechanisms (i.e., sound waves propagating in the photon-baryon fluid) acting 
on scales of order the horizon size at the epoch of decoupling. A similar effect occurs for the CMB 
polarization power spectrum, though in this case it is the horizon size of the re-scattering surface, not the 
`primary' scattering surface, which is imprinted in the observed power spectrum.

Following Peebles (1993), we expect that the observed CMB polarization angular correlation scale will be: 
$\Theta_{{\rm ri}} \sim 0.1 (\Omega_B \Omega h)^{1/3}$ rad. For $\Omega =0.1, \Omega_B = 0.1, h = 
1$ we find $\Theta_{{\rm ri}} \sim 1\arcdeg$, and for $\Omega = 1, \Omega_B = 0.05, h = 0.65$ we find 
$\Theta_{{\rm ri}} \sim 2\arcdeg$.  This new angular scale, absent in non-reionized models, is manifested 
in the spatial polarization correlation function and creates a peak in the reionized polarization power 
spectra at $\ell \leq 20$.

Using CMBFAST, we have generated polarization spectra created by scalar perturbations in a CDM 
dominated, completely reionized, universe with  $ x = 1, \Omega = 1, \Omega_B = 0.05, h = 0.65$. By 
varying the redshift of reionization  in the range $0 < z_{{\rm ri}} < 105$, we compute multiple 
polarization power spectra, which are displayed in figure \ref{figthy2}. The power spectra illustrate the 
main features expected from the theoretical principles detailed above.  Large angular scales correspond to 
modes with wavelengths greater than the width of the last scattering surface.  Prior to recombination 
photons and baryons were tightly coupled and the relatively short timescale for acoustic oscillations 
prevented the formation of long-wavelength perturbations. These effects are particularly evident in models 
without reionization. 

In models with early reionization, polarization at large angular scales is enhanced due to multiple photon 
scatterings following reionization.  At smaller angular scales ($\ell \sim 100$), in models with and without 
reionization, the polarization power spectra exhibit oscillatory behavior, caused by the same type of 
acoustic oscillations which generate the Doppler peaks in the anisotropy power spectra (Frewin, Polnarev, 
\& Coles 1994; Zaldarriaga 1997)\markcite{frew}\markcite{crit}.  Though not relevant for the large 
angular scale considerations discussed here, for $\ell \gg 100$ the polarization is highly suppressed due to 
Silk Damping (Hu 1996).

\placefigure{figthy2}

The power spectra are, effectively, 
predictions of the polarization 
which should be observable given a 
particular observing strategy. 
We will show in section \ref{secobs} 
that the rms polarization expected from 
the spectra shown in Figure 2 
with $\Theta_A = 7\arcdeg$,
 is in the range $0.05 \mu$K$ < I_{{\rm pol}} 
< 1.0\mu$K, where the 
lower limit is standard recombination with no
 reionization, and the upper limit is 
for total reionization
 starting at $z = 105$. These
 limits agree well with
 the analytic estimates for non-standard 
ionization histories discussed in 
subsection 2.1. 
For a $6\arcdeg$ experiment 
and a non-standard ionization history, 
figure 1 predicts a 
polarization level 
of $5 \times 10^{-7}
 \sim 1\mu$K which
 agrees well with our
 numerical simulations of early 
reionization (e.g., for $z_{reionization} 
= 105$). 

\section{Experimental Overview}
\label{secexpt}
Measurement of polarization of the CMB poses a wide variety of experimental  challenges, many of which 
are familiar from the experiments now measuring spatial anisotropy in the CMB.  We describe below the 
design of  POLAR to illustrate the experimental issues that must be addressed in any CMB  polarization 
observation.

POLAR will measure polarization on $7\arcdeg$ scales with two separate radiometers, one in the $K_a$  
frequency band, and one in the $Q$ band, covering the spectrum between 26 and 46 GHz.  These 
radiometers operate simultaneously, and their frequency bands are multiplexed into several sub-bands to 
allow for discrimination against foreground sources. Each radiometer executes a drift scan of the zenith 
with a separate FWHM=$7\arcdeg$ beam produced by a corrugated feed horn antenna. POLAR will 
observe $\sim 36$ different pixels for many months to reach the level of a few $\mu$K per pixel. The 
design builds on techniques developed in previous searches for CMB polarization (Nanos 1979; Lubin 
1980; Lubin \& Smoot 1981; Lubin, Melese, \& Smoot 1983; Wollack et al. 1993; Netterfield et al. 1995) 
\markcite{lube}and is driven by the key issues identified in this paper:  the size and angular scale of the 
anticipated CMB signals, spectral removal of foreground sources, optimization of the observing scheme, 
and anticipated systematic effects.    

\subsection{The Polarimeter}    
Radiation from the sky couples into a corrugated  circular horn antenna (See Fig. \ref{figexpt1}).   This 
antenna  has extremely low sidelobes, near $-80$ dB at $90 \arcdeg$ off axis, in both polarizations, across 
a full waveguide band.  The main lobe of the antenna is gaussian with a FWHM of $7 \arcdeg$ which is 
near the minimum that can easily be obtained without additional optical components such as lenses or 
primary reflectors.  The antenna output couples to an ortho-mode transducer (OMT), a waveguide device 
that decomposes the incoming wave into two orthogonal linear polarization components.  The OMT 
defines the x-y coordinate system of the antenna.

\placefigure{figexpt1}
    
The $Q$ and $U$ Stokes parameters are defined in terms of a coordinate system fixed to the sky.  There 
are several approaches to measuring $Q$ and $U$ for a particular pixel on the sky.  Lubin \& Smoot 
(1981) employ a Dicke switch which alternately couples each of the polarization components from the 
OMT to a low-noise amplifier and square-law detector (Lubin 1980; Lubin \& Smoot 1981).  Phase-
sensitive detection at the modulation frequency of the switch yields the difference between these two 
components, the $Q$ Stokes parameter, and helps overcome $1/f$ noise from the amplifier. One can show 
that after a $45\arcdeg$ rotation about the antenna symmetry axis the instrument measures the $U$ 
parameter. A second technique couples the output of an OMT directly to two square-law detectors 
(Netterfield et al. 1995; Wollack et al. 1997). The beam is switched on the sky to measure the spatial 
anisotropy in two orthogonal polarizations. This approach measures the anisotropy in the $Q$ Stokes 
parameters of the incident radiation field, and provides the most stringent upper limits on the spatial 
anisotropy of the polarization of the CMB.

An alternate approach, employed in POLAR, is the correlation radiometer (Fujimoto 1964; Rohlfs 
1990)\markcite{fuji}\markcite{rohlfs}.   In this instrument the two polarization components are amplified 
in separate parallel amplifier chains;  the output signals are correlated, resulting in an IF signal 
proportional to the $U$ Stokes parameter. This type of  instrument effectively ``chops" between the two 
input RF signals at a frequency which is comparable to that of the RF signals themselves. The correlation 
polarimeter has a $\sqrt 2$ noise advantage over the Dicke-switched approach. An advantage of this 
differencing mechanism is that it has no magnetic or moving parts which have traditionally complicated 
experiments of this type.  After a $45\arcdeg$ rotation the correlator gives the $Q$ parameter.  POLAR 
rotates about the vertical in $45\arcdeg$ steps at a few rpm.  The rotation modulates the output 
sinusoidally between $U$ and $Q$ at twice the rotation frequency and allows the removal of an 
instrumental offset and other instrumental effects that are not modulated at this frequency.

The sensitivity of the polarimeter is determined by low-noise HEMT amplifiers cooled to  15K by  a 
commercial cryocooler.  State-of-the art versions of these amplifiers achieve noise temperatures of  $\sim 
10$K over a bandwidth of $\sim 10$ GHz in both the $K_a$ and $Q$ bands (Pospieszalski 1992; 
Pospieszalski 1995).  This noise temperature is comparable to the antenna temperature of the atmosphere 
at a good observing site.  Nevertheless, long integration periods are required to reach a sensitivity level 
$\simeq 1\mu$K.   The rms noise in a measurement of either $Q$ or $U$ (in antenna temperature) is given 
by the radiometer equation (Krauss, 1982)\markcite{krauss}. For the $Q$ Stokes parameter:   
\be    
\Delta Q_{rms} = \frac{\kappa (T_{rec} + T_{atm} +  
2.7)}{\sqrt{\Delta\nu\tau/2}},
\ee
where $T_{rec}$ and $T_{atm}$ are the receiver and atmospheric noise temperatures, respectively. 
$\tau$ is the total time spent observing the CMB; the time spent either $Q$ or $U$ is $\tau/2$. $\Delta\nu$ 
is the RF bandwidth and $\kappa = \sqrt{2}$ for a correlation radiometer. For the $K_a$ polarimeter, 
$T_{rec} \simeq 20$K and $T_{atm} \simeq 10$K, and $\Delta\nu$ is 10 GHz, resulting in a sensitivity to 
$Q$ or $U$ of  $NET = \Delta Q_{rms}\sqrt{\tau/2} \simeq 460 \mu$K$s^{1/2}$.
For the total polarized intensity we have: $I_{{\rm pol}} = \sqrt{Q^2 + U^2}$. The error in $I_{{\rm 
pol}}$ is $\Delta I_{{\rm pol}} = \sqrt{2}\Delta Q_{rms} = 650  \mu$K$s^{1/2}$, before foreground 
subtraction. To reach a signal level of $ \Delta  I_{{\rm pol}}=  1\mu$K for a single pixel requires an 
integration time of $\sim 120$ hours. Consequently, to measure polarization at a signal-to-noise ratio of 1 
per pixel for all 36 pixels, demands a total observation time of  $\sim 180$ days. The $Q$-band radiometer 
requires a similar amount of time.

\subsection{Systematic Effects} 
Because the anticipated polarization signal is a factor of $\sim 10$ times smaller than the temperature 
anisotropy currently being detected, the understanding of systematic errors is crucial.  Polarization 
experiments have several advantages, however, that promise to make this effort possible.  First, the 
atmosphere is known to be polarized only at a very low level, far below the expected level of CMB 
polarization (see section 4). Additionally, POLAR is essentially a total (polarized) power radiometer, 
which eliminates the comparison of pixels through different airmasses, and at different times. In 
anisotropy observations, beam switching often adds noise and additional chop-dependent signals. 
Potentially, atmospheric effects will have a smaller contribution to this type of experiment than to ground-
based CMB anisotropy experiments and will allow longer observation times than have been possible in the 
past.  Long-term observations are key to understanding and removing systematic effects (Wilkinson 1995; 
Kogut et  al. 1996b; Bennett et al. 1993).  Many spurious instrumental effects can be isolated from 
astrophysical effects by long-term integration tests with the horn antenna replaced by a cold termination.

In Table 2 we list some important systematic effects encountered in previous polarization measurements 
and summarize the solution adopted by POLAR. A full analysis of all potential effects is obviously beyond 
the scope of this paper. In section 4 we discuss in detail the discrimination against foreground sources, 
such as: extragalactic sources, galactic sources, and the atmosphere.   

\section{Foreground Sources}
\label{secfor}
A fundamental question for any measurement of the polarization of the CMB is whether the expected 
signal can be distinguished from polarized foreground sources. While astrophysical (non-cosmological) 
sources of polarized radiation are of interest for other fields, the measurement of CMB polarization is our 
main objective, so these sources are spurious effects. These foreground sources all have spectra that are 
distinct from that of the CMB, and in principle can be distinguished  from it by multi-frequency 
measurements.  This technique has been employed for observations of CMB anisotropy (Brandt et al. 
1994).  However, polarized foreground spectra have not been studied as extensively. To estimate the 
intensity and spectra of these foreground sources, we rely on theoretical predictions and extrapolations 
from measurements at different frequencies of the antenna temperatures of these foregrounds. Here we 
summarize the properties of atmospheric and astrophysical (though non-cosmological) foreground sources.

{\it Synchrotron Emission.}
Diffuse galactic synchrotron radiation arises from ionized regions of our galaxy that posses magnetic 
fields. The antenna temperature of synchrotron emission obeys a power law: 
\[T_{synchrotron}(\nu)\propto \nu^{\beta},\]
where $\beta$ is referred to as the {\it synchrotron spectral index}. The polarization level $\Pi$ of 
synchrotron radiation is related to the spectral index (Cortiglioni \& Spoelstra 1995)\markcite{cortiglioni}:
\[ \Pi = \frac{3\beta+3}{3\beta+1} \]. 
Faraday rotation and non-uniform magnetic fields will reduce the level of polarization given by this 
equation. Full-sky polarization maps made at radio frequencies (Brouw and Spoelstra 1976) have been 
extrapolated to millimeter-wave frequencies assuming a power-law spectrum (Lubin \& Smoot 
1981)\markcite{lub}. The radiation is linearly polarized between approximately 10\% and 75\%, 
depending on galactic coordinates.   Below 80 GHz the polarized synchrotron emission dominates all 
sources, including the CMB if it is polarized at the $1\times 10^{-6}$ level, as shown in Figure 
\ref{figforg1}.  We estimate the spectrum of the synchrotron radiation by extrapolating the Brouw \& 
Spoelstra (1976) \markcite{brouw} measurement at 1411 MHz to millimeter wavelengths with the 
modified power-law spectrum used to fit the {\em COBE} DMR data (Bennett et al. 1992).  For our 
modeling purposes we choose: $-2.9 \geq \beta \geq -3.2$ total intensity of approximately 50 $\mu$K at 30 
GHz, typical of high galactic latitudes (Kogut et al. 1996a; Bennett et al. 
1992)\markcite{kog1}\markcite{ben1}, and $\Pi$= 75$\%$. \markcite{brouw}.

{\it Bremsstrahlung Emission.}
Bremsstrahlung, or free-free, emission from ionized hydrogen (HII) regions is not polarized (Rybicki and 
Lightman 1979)\markcite{ryb}. However, bremsstrahlung emission will be polarized via Thomson 
scattering by the electrons in the H II region itself.  The rescattered radiation will be polarized tangentially 
to the edges of the cloud, at a maximum level of approximately 10\% for an optically thick cloud.  The 
locations and emissivities of galactic H II regions are not well known, but Bennett et al. (1992) model the 
bremsstrahlung emission in the galaxy by subtracting a synchrotron model from microwave sky maps.  
In any case the polarization in the rescattered bremsstrahlung emission will be at least an order of 
magnitude smaller than the polarized synchrotron signal at frequencies greater than 10 GHz. We quote the 
result of Bennett et al. (1992) \markcite{ben1} that 
\begin{equation}
T_{bremsstrahlung}\propto\nu^{-2.15}\label{eq:sb},
\ee
with total intensity  $\sim 40 \mu$K at 30 GHz 
(Kogut et al. 1996a; Bennett et al. 1992).

{\it Dust Emission.}
The polarization level of interstellar dust is not well known.  At low galactic latitudes thermal emission 
from dust particles dominates the near infrared spectrum.  Depending on the shape and the alignment of 
dust particles, emission from dust particles may be highly polarized  (Wright 1987)\markcite{wri}.  Using 
the dust spectrum measured by the {\em COBE} FIRAS (Wright et al. 1991) normalized to the IRAS 100 
micron map, we find that high galactic latitude dust emission is negligible below 80 GHz, even when 
assumed to be 100\% polarized. We use the two temperature dust model (Wright et al. 1991) :
\begin{equation}
T_{dust} \propto \frac{c^2}{2\nu^2k} \left( \frac{\nu}{900 {\rm GHz}}
\right)^2\left[B_{\nu}(20.4{\rm K}) + 6.7 B_{\nu}(4.77 {\rm K})\right].
\label{eq:sd} 
\end{equation}
At high galactic latitudes $T_{dust} \sim 10\mu$K at 200 GHz (Kogut et al. 1996a; Bennett et al. 
1992)\markcite{ben1}.

{\it Extragalactic Point Sources.}
The dominant radiation mechanism for extragalactic radio sources is synchrotron emission (Saikia \& 
Salter 1988).  These sources have a net polarization of $<  20\%$.  Calculations made by Franceschini et 
al. (1989)\markcite{franc} of the temperature fluctuations in measurements of anisotropy of the CMB 
arising from unresolved, randomly distributed sources show that they contribute negligibly at 30 GHz to a 
$7\arcdeg$  anisotropy experiment. If the orientations of the polarization vectors of these sources are 
uncorrelated over $7\arcdeg$ regions, we would also expect a negligible contribution to the signal 
observed POLAR. We ignore the contribution of these sources in our foreground modeling. 

{\it Atmospheric emission}. The antenna temperature of the earth's atmosphere between 10 and 60 GHz is 
dominated by an emission feature at $\sim 22$ GHz caused by atmospheric water vapor, and a series of 
emission lines at $\sim 60$ GHz due to molecular oxygen. In the absence of external fields, neither of 
these atmospheric components is known to emit polarized radiation in the frequency range of interest. 
However, Zeeman splitting of the energy levels of atmospheric molecules by the magnetic field of the 
earth can produce polarized emission. The valence band of water is completely full, and thus, does not 
exhibit Zeeman splitting. However,  the $O_2$ molecule has a non-zero magnetic moment due to its two 
unpaired valence electrons which interact with the Earth's magnetic field. We here discuss polarized 
emission from mesospheric oxygen, and show that it is negligible in comparison with the expected 
polarized intensity of the CMB. 

The Zeeman effect breaks the energy degeneracy of the two unpaired valence electrons of molecular 
oxygen. The total angular momentum quantum number of the oxygen molecule is $j=1$, which implies 
that the oxygen molecule's rotational spectral lines are Zeeman split into $2j + 1 = 3$ distinct lines.  Dipole 
radiation selection rules for transitions between these levels permit transitions as long as the change in 
magnetic quantum number, $m$, is: $\Delta m = 0, \pm 1$. Transitions with $\Delta m = +1$, for example, 
correspond to the absorption of a right circularly polarized photon or the emission of a left circularly 
polarized photon. The absorption and emission properties depend, therefore, on both the frequency and 
polarization of the radiation. The frequency of each Zeeman split level is (Liebe 1981)\markcite{liebe}:

\be
\nu_Z = \nu_0 + 2.803 \times 10^{-3} B \eta (\Delta m) {\rm   [GHz]}\label{eqat1}
\ee
where $\nu_0$ is the unperturbed frequency, $\eta$ is a shift factor with
$|\eta| \leq 1$, and $B$ is the magnitude of the earth's magnetic field,
typically 0.5 Gauss throughout the mesosphere. The largest possible frequency
shifts occur for $\eta = \pm 1$, which imply that the center frequencies for the
polarized emission components will be confined to within $1.4$ MHz of the
unsplit center frequency. In principle, emission at these split frequencies
could be up to 100\% circularly polarized. Away from the center frequencies, the
total intensity of emitted radiation decays with frequency as: $ I \sim
\frac{1}{(\nu - \nu_0)^2}$ (Rosenkranz 1994). For a small shift in frequency,
$\Delta \nu_0$, away from the center frequency, the first order fractional
change in emissivity can be shown to be: \be \frac{\Delta I}{I} = \frac{2\Delta
\nu_0}{\nu - \nu_0}.  \ee For a single Zeeman split component, $\frac{\Delta
I}{I} = \frac{2\Delta \nu_{Z,\Delta m}}{\nu - \nu_Z}$, where $\Delta
\nu_{Z,\Delta m} = \nu_Z - \nu_0 = 2.803 \times 10^{-3} B \eta (\Delta m) {\rm
[GHz]}$, from equation \ref{eqat1}. To obtain the total contribution to the
emission of both polarization components we must sum over left-handed and
right-handed contributions: \be \frac{\Delta I_{{\rm tot}}}{I} = \sum_{\Delta m
= \pm 1} \frac{2\Delta \nu_{Z,\Delta m}}{\nu - \nu_Z} \ee

However, we have for the shift factor in equation \ref{eqat1}: $\eta(\Delta m = +1) = - \eta(\Delta m = -1)$ 
so the net effect on the emissivity is exactly canceled out by the two circularly polarized components 
(Rosenkranz 1994).  Any second order contributions to the emission scale as $\sim \frac{1}{(\nu - 
\nu_Z)^2}$, which implies a contribution of $< 10^{-8}$K for $26 \leq \nu \leq 46$ GHz, i.e. the 
frequency band which POLAR will probe (Crill 1995). For these frequencies of observation there is also a 
small Faraday rotation of the plane of polarization of the CMB. Rosenkranz \& Staelin (1988) show that 
the rotation of the plane of polarization will be less than $\sim 10^{-2}$ degrees for these frequencies. 
Therefore, both the polarized emission and Faraday rotation of the atmosphere are negligible effects in the 
range of frequencies which POLAR will probe. 

Of all the relevant foreground sources, only diffuse galactic synchrotron radiation and dust are expected to 
appear at a level comparable to  the anticipated polarized CMB signals, (see Figure \ref{figforg1}).   In the 
next section we discuss techniques to remove spurious  foreground sources using multi-frequency 
observations.

\placefigure{figforg1}

\section{Foreground Removal}
\label{secsub}
Here, we estimate our ability to subtract foreground sources in the presence of atmospheric and 
instrumental noise.  We apply two different approaches to foreground removal. In the first approach we 
compute the anticipated  error in our recovery of $Q$ and $U$ using an analytic technique developed by 
Dodelson (1995). In the second, Monte Carlo, approach we create simulated data and use a least-square 
fitting procedure to recover the underlying CMB $Q$ and $U$ parameters. We evaluate the quality of the 
fit by comparing the recovered and true CMB values.

\subsection{Foreground Modeling}
If there are several sources of polarized radiation at a given frequency, in a given pixel, the Stokes 
parameters describing the total radiation is the sum of the Stokes parameters of each source. Following 
Dodelson (1995), we represent the total signal as:
\be
{\bf Q }= \sum_{i=0}^{3}q^i {\bf F^i} + {\bf N}_q \mbox{ and } 
{\bf U}=  \sum_{i=0}^{3}u^i {\bf F^i} + {\bf N}_u
\ee
where the labels $0,1,2,3$ represent the CMB, synchrotron radiation, bremsstrahlung, and dust emission 
contributions to the total signal, respectively. The $Q$ and $U$ are expressed in terms of antenna 
temperature. The $q^i$ and $u^i$ are the thermodynamic temperatures of each signal source, and the 
${\bf F}^i$ are the spectral shapes of each source, following section \ref{secfor}. $\bf Q$ and $\bf U$ are 
written as vectors\markcite{dod} with dimensions equal to the number of observation frequencies. The 
components of $\bf Q$ and $\bf U$ are the Stokes parameters at that frequency, e.g.: ${\bf Q} = [ 
Q(\nu_1), Q(\nu_2), \cdots , Q(\nu_{N_{obs}})]$. At frequency $\nu$:
\be
Q(\nu)= q^0 \frac{c^2}{2\nu^2}\frac{{\rm d} B_{\nu}(T=2.728{\rm K})}{{\rm d}T}+ q^1 \nu^{\beta} 
+ q^2\nu^{-2.15} \\ 
+ q^3  \frac{c^2}{2\nu^2 k}\nu^2(B_{\nu}(20.4{\rm K})+6.7 B_{\nu}(4.77{\rm K})) + N_q(\nu)
\label{eq:tot}
\ee

A similar equation holds for $U$.  $B_{\nu}(T)$ is a blackbody spectrum with thermodynamic 
temperature $T$. ${\bf N}_q$ and ${\bf N}_u$
each represent the combined contributions from instrument and atmospheric noise.
For simplicity, we assume that the noise has the same Gaussian distribution for both $Q$ and $U$, and is 
the same for each frequency channel. Furthermore, $\langle N \rangle = 0 $ and $\langle N^2 
\rangle=\sigma^2.$

The unknowns in the signal are the coefficients $q^i$ and $u^i$ and the synchrotron spectral index, 
$\beta$. To simplify our notation, we write these intensity coefficients as vectors: ${\bf q} = [q^0, q^1, 
q^2, q^3].$ Note that $q^0$ is the thermodynamic temperature corresponding to the $Q$ Stokes parameter 
of the CMB, and is frequency independent.

\subsection{Error Analysis: Analytic Method}

Our goal is to recover ${\bf q}$ and ${\bf u}$, particularly $q^0$ and $u^0$ from measured values of the 
$Q$ and $U$ Stokes parameters for each pixel at multiple observation frequencies. Given $Q$ and $U$ it 
is convenient to consider $I_{{\rm pol}}$, the linearly polarized intensity of the radiation $I_{{\rm 
pol}}=\sqrt{Q^2+U^2}$,
and the polar angle $|\alpha|=\frac{1}{2}\tan^{-1}\left(\frac{U}{Q}\right)$. Following Dodelson (1995), 
we wish to minimize the difference between the true CMB polarization signal and our best estimate of it. 
The fitting process is a $\chi^2$ minimization for the linear unknowns 
which produces our best estimates for the intensity coefficients:
\[{\bf q'}={\bf \kappa^{-1}} {\bf \eta}_q 
\mbox{ and }
{\bf u'}={\bf \kappa^{-1}} {\bf \eta}_u \]
where 
\[\kappa^{ij}=\frac{{\bf F}^i \cdot {\bf F}^j }{\sigma^2} \mbox{ and } \eta_q^i =\frac{ {\bf Q}\cdot {\bf 
F}^i}{\sigma^2} \mbox{, } \eta_u^i =\frac{ {\bf U}\cdot {\bf F}^i}{\sigma^2}.\]

The ${\bf F}^i$ encode only the spectral shape (frequency dependence) of each foreground component. 
They are defined to be unit vectors such that ${\bf F}^i \cdot {\bf F}^j = 1$ implies that ${\bf F}^i $ and 
${\bf F}^j$ have the same frequency dependence. Note that the dimension of ${\bf q'}$ and ${\bf u'}$ 
may be different from that of $\bf q$ and $\bf u$, if we choose to restrict our attention to fewer 
foregrounds. The upper limit on the dimension of ${\bf q'}$ and ${\bf u'}$ is the number of frequencies of 
observation. 

We define 
\[i_{{\rm pol}}'^{ i} = \sqrt{(q'^{i})^2 + (u'^{i})^2} \mbox{ and }|\alpha '^i| = \frac{1}{2} \tan^{-
1}\left(\frac{u'^i}{q'^i}\right),\]
as the $i$th component's polarized intensity coefficient and polar angle. The error in our analytic fitting 
process is:
\be
 {\rm error}(i_{{\rm pol}}'^i) = \sqrt{(\kappa^{-1})^{ii}} \equiv \sigma \epsilon^i 
\label{eqjspi}
\ee
and
\be 
{\rm error}(|\alpha'^i|)=\frac{1}{2}\frac{\sqrt{(\kappa^{-1})^{ii}}}{i_{{\rm pol}}'^i}
\equiv \frac{1}{2}\frac{\sigma \epsilon^i}{i_{{\rm pol}}'^i}.
\label{eqjsal}
\ee

We are interested in the recovery of the true CMB polarization intensity and orientation angle in the 
presence of the foregrounds. The error in the fitted CMB polarization intensity coefficient, ${\rm 
error}(i_{{\rm pol}}'^0)$, is found from equation (\ref{eqjspi}) to be the standard deviation of the system 
(instrumental + atmospheric) noise multiplied by a factor, $\epsilon^0$,  which depends only on the 
frequencies of observation and the choice of foregrounds for which one fits. It can be normalized such that 
its minimum value (i.e., with no foregrounds or system noise) is 1.0. The normalized parameter, known as 
the {\it Foreground Degradation Factor} (Dodelson 1995), is
\[{\rm FDF} = \epsilon^0 \sqrt{{\bf F^0}\cdot{\bf F^0}} .\] 

An optimized experiment will minimize not only the system noise $\sigma$, but also the FDF. Another 
contribution to the analytic error defined by Dodelson (1995) is the error arising from the uncertainty in 
the spectral shape functions,
${\bf F}^i$, of the foregrounds, which vary with position on the sky. The situation is now more 
complicated because we must compute these contributions separately for both $Q$ and $U$. This error, 
called $\sigma_{{\rm shape}}$ by Dodelson (1995), is negligible for our most important foreground, 
synchrotron radiation, even though we only know $\beta$ to 10\%.

\subsection{Error Analysis: Monte Carlo Method}
To corroborate the results of the analytic error calculation, we perform a more
explicit
foreground removal simulation.  The approach is similar to that of Brandt et
al. (1994)\markcite{bran}.  First we
choose a particular  set of observing frequencies.  Using equation (\ref{eq:tot}),  at  each
pixel we
create simulated signals for each frequency.  Using foregrounds levels typical of
high-galactic
latitude regions.  While the foreground temperatures are fairly well known, the
polarization levels are not and we choose the most conservative estimates:
$75\%$ for
synchrotron, $10\%$ for bremsstrahlung, and $100\%$ for dust.  The CMB signal
is
chosen nominally at $10\%$ of the anisotropy level ($\sim 3 \mu$K), but the
results of the
simulation are independent of the exact value.  For each of these signals we
keep the true intensity, $\equiv {\rm i_{pol}}^i$, fixed but allow the orientation, $\equiv \alpha^i$, to
vary randomly. Noise is chosen from a gaussian distribution with standard
deviation $\sigma$ in each
frequency channel and added as in equation (\ref{eq:tot}).  We then perform a least-squares
fit to a particular signal and foreground model and determine our best-fit
values for the recovered CMB intensity ${\rm i_{pol}}'^0$ and orientation angle $\alpha'^0$. We repeat 
this
process 200 times, generating
 new values for the system noise and the $\alpha^i$ each time.  Finally, we compute
the rms of the difference between the 200 input CMB coefficients, ${\rm i_{pol}}^0$, and
the corresponding best-fit CMB values, $i_{{\rm pol}}'^0$. A similar procedure is conducted for the true 
CMB polarization orientation angle, $\alpha^0$, and the recovered angle, $\alpha'^0.$  In this process we 
average over many different relative orientations of the
polarized foreground vectors.
 
Typically the number of foregrounds (e.g., 4) simulated is greater than the number of frequency channels 
(e.g., 2 or 3), so we can fit for only some of the signals (e.g., the CMB and one or two foregrounds).   As
shown below, for a judicious choice of the observation frequency channels, the neglected foregrounds will 
not
contribute significantly to the error in the recovery of the CMB signal.

The primary unknown which is not determined by this analysis is the synchrotron
spectral index, which varies with sky position. We estimate an uncertainty in
the synchrotron spectral index by using a range of values for the fitting
function $F^1(\nu)=\nu^{\beta}$. We evaluate of $F^1(\nu)$ for three values of
the spectral index: $\beta= -2.9, -3.05, -3.2$.

\subsection{Results of Error Analysis}
Here we concentrate on a system noise range of 0.2 to 50 $\mu$K (after a long integration on each pixel), 
and show the results of a two channel configuration observing at 30 and 40 GHz, which is appropriate for 
POLAR. Here, the only foreground considered is synchrotron radiation, and we find that the degradation 
to our experimental sensitivity is: FDF$= 2.7$.  Other frequency channel configurations show, very 
generally, that in order to fit for an extra foreground source (e.g. dust emission)
without
a severe increase in the FDF, one must observe more than three frequency channels.

The plots of polarized intensity coefficient error versus system noise demonstrate that the quality of the 
recovery is fairly insensitive to
 the value of the input 
synchrotron spectral index. 
The Monte Carlo results are plotted as 
points, and the analytic result is plotted as a line. 
The excellent agreement of our analytic and Monte Carlo results
confirms the idea that a successful observation should have a small 
$\epsilon^0$, or equivalently, a low FDF.

\placefigure{figsub1}

For CMB polar angle recovery, the plotted symbols represent Monte Carlo
simulations at the two extreme values of synchrotron spectral index.
Generally, the polar angle is more difficult to 
determine than the total linear polarized intensity, even at very low  
values of system noise. However, frequency configurations which yield an accurate recovery of the total 
polarized intensity will often give a more accurate
recovery of the polar angle as well.

\placefigure{figsub2}

From our analysis and simulation of a CMB polarization, we conclude that for a dual frequency 
experiment observing frequencies below 60 GHz, the most
important challenge is to reduce the system noise. The contribution of the system noise is far more 
important than that of combining 
bremsstrahlung
and synchrotron in the fitting process or reducing the uncertainty in the synchrotron 
spectral
index used to fit the data.
We conclude that an experiment of the 30-40 GHz type with system noise lower than
1 $\mu$K is capable of discriminating a $\sim 1- 3 \mu$K CMB polarization signal from polarized galactic 
synchrotron radiation.

 \section{Observation Strategy}
\label{secobs}
Constraints on, or detection of, the polarization of the CMB and its associated power spectrum depend 
greatly on the amount of sky coverage of the observation and the sensitivity of the radiometer. Sensitivity 
considerations are common to all CMB observations: time limitations restrict signal integration, and 
constrain the amount of sky coverage. We must reach a compromise between the integration time required 
to achieve the desired signal-to-noise ratio while also sampling a representative distribution of celestial 
regions. We now discuss our observing strategy in the context of the achievable level of sensitivity of 
POLAR.

\subsection{Sky Coverage}
 
The trade-off between the signal-to-noise ratio per pixel, s:n, and the total number of pixels arises 
frequently in designing CMB observing strategies. The ultimate goal is to discriminate between two 
hypotheses: ($H_0$ and $H_{{\rm pol}}$), which are the null hypothesis of an unpolarized CMB, and the 
hypothesis that the CMB is polarized at a particular level. This discrimination is quantified by the 
confidence level and power of the measurement. POLAR seeks to make a primary detection, namely, a 
detection of a signal for which no prior detections have been made. There are four possible outcomes: the 
first two are correct detections of either $H_0$ or $H_{{\rm pol}}$. The other two outcomes are 
erroneous detections: a false positive or a false negative detection. Maximizing the confidence level 
minimizes the probability of a false detection, while maximizing the power minimizes the probability of 
making a false negative detection. Power quantifies the ability of an experiment to distinguish between 
competing hypotheses, and confidence quantifies the certainty of the detection. Traditionally, most CMB 
experiments have quoted only confidence intervals.  Keating \& Polnarev (1997) argue that the confidence 
level should be equal to the discriminating power for a primary detection.

For a fixed total amount of observing time we wish to find the number of pixels which maximizes the 
confidence level and the power of the experiment. We estimate the polarized CMB intensity from models 
described above. While our s:n per pixel is less than one (see section \ref{secexpt}), we expect that our 
signal-to-noise ratio defined over the total observation time, denoted by $R$, will be $R \sim 1$. For 
$R=1$ it is shown that the optimum number of pixels is $\sim 10$, with a very weak dependence of the 
optimum on $N$. In order to minimize the susceptibility to certain systematic effects, as discussed above, 
POLAR observes a constant declination which corresponds to the zenith. The number of independent 
pixels which will be observed is $N = 360\arcdeg \cos\theta_{{\rm lat}}/\Theta_A$, where $\theta_{{\rm 
lat}}$ is the latitude of Madison, WI and $\Theta_A$ is the  antenna full width at half maximum, 
$FWHM$. We find $N \sim 36$, though we are free to pixelize our data into bins which are smaller than 
this, for instance pixelizing at the `Gaussian width' of our antenna, $\sigma_B = FWHM/2\sqrt{\ln2}$. 
Observing fewer pixels would increase the per-pixel s:n, but would either require tracking individual pixels 
over large angles on the sky or tilting the radiometer toward the north celestial pole. Either approach would 
undoubtedly introduce gravitationally modulated systematic effects into our data. Additionally, with $N 
\sim 36$ we will have good coverage of the galaxy, which should allow for both the removal of 
foregrounds and a comparison with previous galactic polarization surveys (e.g., Brouw \& Spoelstra 1976). 
For $N=36$ and $R=1$, POLAR can expect to make a detection at the $\sim 55\%$ level of confidence 
and power for a CMB which is polarized at $ \sim 1 \mu$K. Of course, it is still possible to quote results 
with confidence arbitrarily close to $100\%$, but this would be at the expense of ability to reject the null-
hypothesis (i.e., low power). In the context of a primary detection, requiring equal levels of confidence and 
power seems the most reasonable compromise.

\subsection{Sensitivity to the Power Spectrum}

The observation's sensitivity to CMB fluctuations is quantified by its window function, $W_{\ell}$. The 
observed two-point correlation function is related to the power spectrum and window function as  follows:
\be
\lan Q({\bf \hat{n}_1})Q({\bf \hat{n}_2}) + U({\bf \hat{n}_1}) U({\bf \hat{n}_2}) \ran = 
\frac{1}{4\pi}\sum_{l=0}^{\infty}(2\ell + 1)\times C_{\ell}^{\Pi}W_{\ell}^{\theta} \times 
P_{\ell}(\cos\theta)
,\label{eqpolpol}
\ee
where, for example, $Q({\bf \hat{n}})$ is the Stokes parameter measured for a pixel located in the 
direction ${\bf \hat{n}}$. $C_{\ell}^{\Pi}$ is the power spectrum describing the degree of polarization on 
angular scales characterized by $\ell$, $W_{\ell}^{\theta}$ is the window function of this observing 
scheme, $\cos(\theta)= {\bf \hat{n}_1} \cdot {\bf \hat{n}_2}$ is the separation between pixels in this 
observing scheme.

The analysis differs from that of most anisotropy experiments in several respects.  The primary difference 
is that the observations are total-power in nature, rather than differential. The window functions for this 
experiment will reflect the fact that there is no  ``chopping'' of the beam in sky-position inherent in the 
observation.  Single pixels will be formed by binning the acquired data, and differencing between pixels 
can be performed during analysis of the data; not during acquisition. This approach avoids systematic 
effects which can arise from mechanical chopping mechanisms. Data from POLAR will be analyzed using 
a variety of synthesized window functions, each sensitive to a different angular scale.  In this respect the 
analysis will be similar to that of the Saskatoon Big Plate observations (Netterfield et al. 1995, Wollack et 
al. 1997)\markcite{net}.

Window functions for observations with less than full-sky coverage are specified by three functions: the 
beam profile function, the beam position function, and the weighting or `lock-in' function (White \& 
Srednicki 1995)\markcite{whitesred}. The beam profile function, $G(\theta,\theta_i,\sigma_B)$, where 
$\sigma_B$ is the beamwidth, quantifies the directional response of the antenna, which is assumed to be 
gaussian.  $G(\theta,\theta_i,\sigma_B)$ effectively samples all angular scales larger than, approximately, 
the angular size of the beam. The angular coordinates of the center of the beam are specified by the beam 
position function, $\theta_i$, and the lock-in function, $w_i^{\alpha}$, is the weighting of each of the $N$ 
binned pixels indexed by $i$, for the scan strategy denoted by $\alpha$. We have 
$G(\theta,\theta_i,\sigma_B) = \frac{1}{2\pi \sigma_B^2} \exp\left( -\frac{(\theta - 
\theta_i)^2}{2\sigma_B^2} \right)$, where $\sigma_B = FWHM/2\sqrt{2\ln2} = 0.052$.

Following  (White \& Srednicki 1995),  the window functions are:
\be
W^{\alpha\beta}_{\ell} 
\equiv \int d{\bf \hat{x}_1} \int d{\bf \hat{x}_2} 
H^{\alpha} ({\bf  \hat{x}_1}) H^{\beta} ({\bf \hat{x}_2}) P_{\ell} ({\bf \hat{x}_1} \cdot {\bf 
\hat{x}_1})
\label{eqwin}
\ee
where $P_{\ell}$ are the Legendre polynomials, and $H^{\alpha}({\bf \hat{x}} = \sum_i w_i^{\alpha} 
G(\theta, \theta_i,\sigma_B)$ quantifies the response of the antenna (for a differencing strategy indexed by 
$\alpha$), when pointed in the direction of  ${\bf \hat{x}}$. For a gaussian beam:  $H^{\alpha}({\bf 
\hat{x}}) = \sum_i w_i^{\alpha} \frac{1}{2\pi \sigma_B^2} \exp\left( -\frac{(\theta - 
\theta_i)^2}{2\sigma_B^2} \right).$

By varying the weight function we will obtain window functions ``tuned'' to sample specific multipole-
space regions. For example, $w_i^{\alpha} = (-1)^{i+1}$, differences pairs of nearest neighbor pixels. 
Pairs of pixels will be separated by a constant angle for each differencing strategy indexed by $\alpha$, 
which runs from 1 to 18, corresponding to the $N/2$  distinct two-pixel difference window functions of 
POLAR.  The number of unique two-pixel combinations, $k$, is given by $k = \frac{N!}{(N - m)!m!}$, 
where $N$ is the number of binned pixels, and $m$ is the number of beams which are differenced.  For 
$N=36$ pixels, differenced pairwise ($m = 2$), $k = 630$.  We note that the total power nature of 
POLAR, as well as the insensitivity of our experiment to atmospheric emission and thermal gradients, will 
allow us to perform two-pixel differencing, in contrast to the majority of ground based CMB anisotropy 
observations, which typically difference three or more pixels.

To estimate the rms polarization for the $FWHM = 7\arcdeg$ polarimeter with 36 pixels, we have 
calculated the single pixel window function of POLAR, $W^{\alpha\alpha}_{\ell}$.  This contains only 
the diagonal elements of the window function matrix from  (\ref{eqwin}). This quantity will allow us to 
determine the relationship between the measured pixels and the underlying spectrum which is responsible 
for a particular realization on the sky. The polarization two-point covariance matrix (at zero-lag) for a 
particular theoretical model is given by:
\be
C^{\alpha\beta}_{\Pi} = \frac{1}{4\pi} \sum_{\ell} (2\ell +1) C_{\ell}^{\Pi}W^{\alpha\beta}_{\ell}
\ee
To compute the theoretical rms amplitude, we extract the square root of the diagonal elements of 
$C^{\alpha\beta}_{\Pi}$ and obtain:
\[ \Pi_{rms}^{\alpha} =\sqrt{\frac{1}{4\pi} \sum_{\ell} (2\ell +1) 
C_{\ell}^{\Pi}W^{\alpha\alpha}_{\ell}},\]
where the $C_{\ell}^{\Pi}$ have been introduced in section  \ref{secthy}. We show this window function 
in Figure \ref{figthy2}. Of course, the off-diagonal components of the window function matrix will contain 
additional information about cross-correlation between pixels, as opposed to $W^{\alpha\alpha}_{\ell}$, 
which is the auto-correlation. The off-diagonal window function matrix elements will increase the effective 
signal-to-noise ratio of the experiment as a whole, and can be `tuned' to sample specific $\ell$-space 
regions, up to the 
cutoff $\ell$ of the antenna, similar to the analysis of (Netterfield et al. 1995). For anisotropy experiments 
it is conventional also to quote the band power, which is independent of the details of the experiment. This 
approach facilitates comparisons between experiments, and contains information equivalent to that of  the 
rms amplitude.

\subsection{Estimated Signal Level and Uncertainty}
 
The primary goal of POLAR is to measure the polarization of the CMB. We have shown in section 2, that 
the level of polarization is extremely sensitive to the ionization history of the universe, both before and 
after recombination. We expect, then, that the observed polarization signal will depend critically on the 
optical depth, $\tau$, for photons back to the last scattering surface. A preliminary estimate of the effect of 
reionization can be obtained by computing the expected rms polarization and associated experimental 
uncertainty for models of a reionized universe. We will now compute the effect of reionization on the 
power spectrum, and demonstrate that the characteristic signature of an early reionization is, in principle, 
detectable by POLAR.

Figure \ref{figthy2} displays ${C^{\Pi}_{\ell}}$, for the power spectrum computed using CMBFAST 
(Seljak \& Zaldarriaga 1996a) for various totally reionized (ionized fraction $x = 1$) scenarios, 
parameterized by the redshift of reionization $z_{{\rm ri}}$. In Figure \ref{figobs1} we plot the expected 
rms polarization vs. $z_{{\rm ri}}$, for $0 <  z_{{\rm ri}} < 105$, along with the statistical $1\sigma$ 
uncertainties we expect based on our NET and observation time. The underlying power spectrum is a 
generic CDM model with $\Omega = 1, \Omega_B = 0.05, h = 0.65,\Lambda = 0$, and pure scalar 
perturbations. The inclusion of a tensor component should enhance the large angular scale polarization 
(Crittenden, Davis, \& Steinhardt 1993; Crittenden, Coulson, \& Turok 1995), so this figure 
underestimates the rms polarization predicted by some cosmological models. This figure suggests that 
POLAR could begin to detect polarization of the CMB at the $1\sigma$ level if the universe became 
completely reionized at a redshift  $z_{{\rm ri}} > 45$. 
\placefigure{figobs1}

\section{Related Topics}
\label{topics}
Although no detections of the polarization have been made, we can glean information about CMB 
polarization from anisotropy detections. In principle, these detections can be utilized to refine a 
polarization observing strategy. Here we briefly discuss the possibility of designing a polarization 
observing strategy, utilizing  information from a well sampled anisotropy map.

Anisotropy and polarization are invariably spatially correlated with one-another, and additional 
cosmological information may be obtained by studying polarization-polarization and polarization-
anisotropy correlation functions. Correlation provides the only link between previous detections (of 
anisotropy), and the proposed measurements discussed in this article. CMB polarization can be 
decomposed into two components: one which is spatially correlated with the temperature anisotropy, and 
another, larger component which is uncorrelated. Ng \& Ng (1996) and Crittenden, Coulson, \& Turok 
(1995), demonstrate that, given a high-resolution CMB temperature map, it would be possible to identify 
celestial regions which are statistically more likely to posses higher levels of the correlated polarization 
component.  As shown in Coulson, Crittenden, \& Turok (1994), the uncorrelated polarization component 
dominates the correlated component by a factor of at least three.  

For detector-noise limited polarization experiments, it can be advantageous to search for polarization-
anisotropy $\lan QT \ran$ correlation in addition to polarization-polarization $\lan QQ \ran$ cross-
correlation. If the noise in the temperature anisotropy map is negligible in comparison with the noise of the 
polarization measurement, $\sigma$, the error in $\lan QT \ran$ will be linear in $\sigma$ while the 
variance in the polarization cross-correlation function grows as $\sigma^2$.  In this limit it becomes 
advantageous to search for correlation.

Power spectrum generation programs, such as CMBFAST, compute $\lan QT \ran$, along with the 
anisotropy and polarization spectra. This allows one to predict the distribution of polarization which is 
correlated with the anisotropy given a well-sampled anisotropy map. Finally, Crittenden, Coulson, \& 
Turok (1995) \markcite{coul} describe potentially observable distributions of correlated polarization 
vectors on the sky which result from the velocity field of the photon-electron plasma and the type of metric 
perturbation which generates the anisotropy. If the angular resolution and experimental sensitivity of 
measurements improve by several orders of magnitude over the current levels, these patterns could prove 
to be direct observables of the microphysical properties of the photon-baryon fluid at the moment of 
decoupling. 

We also briefly discuss an effect which is peculiar to CMB polarization measurements: Faraday 
depolarization of the CMB by primordial magnetic fields. Faraday depolarization causes the plane of CMB 
polarization to rotate differentially due to a residual primordial magnetic field, which may have existed 
during the epoch of recombination (Basko \& Polnarev 1980). The effect is akin to optical dichroism, 
familiar from the polarization of visible light. The net depolarization is frequency dependent, as the 
rotation of the plane of polarization of individual frequency components scales as $\sim 1/\nu^2$. This 
effect is expected to be non-negligible below 30 GHz, for reasonable values of the primordial magnetic 
field (Harrari, Hayward, \& Zaldarriaga 1997; Kosowsky \& Loeb 1996). Detection of this effect is 
unlikely until the polarization of the CMB has been detected over a large frequency bandwidth, and with 
high angular resolution. Neither of the above mentioned phenomena are immediately relevant for 
polarization observations, but are nonetheless quite intriguing. 

\section{Conclusion}
\label{seccon}
We have demonstrated that the detection of the polarization of the CMB is difficult but 
technologically feasible. A detection would permit the discrimination between heretofore degenerate 
theoretical predictions. Polarization of the CMB has a unique signature in both real and Fourier space, as 
well as distinct spectral characteristics. A detection of polarization, in conjunction with the current 
detections of CMB anisotropy could be the best available probe of the ionization history of the pre-galactic 
medium. This epoch of cosmic evolution is of great interest, and supplemental information from 
polarization detection could greatly advance our knowledge of the formation of structure in the early 
universe. The current generation of anisotropy measurements are sufficiently refined that the fundamental 
parameters of classical cosmology are beginning to be determined. Detection of polarization of the CMB 
also promises numerous dividends throughout cosmology, and one readily observes that the status of 
polarization observations today is reminiscent of the status of anisotropy measurements a decade ago. 

\acknowledgments

The authors are indebted to 
many people who have 
worked on the planning and development of  this 
measurement.  Khurram Farooqui and 
Grant Wilson designed an early version of the  instrument and 
observing scheme.  Brendan Crill
 tested the polarimeter
 and calculated
 the atmospheric emission.  Victor 
Derderian designed the rotation 
platform of the instrument. Melvin Phua and Nathan Stebor helped to 
diagnose systematic effects. 
 Dave Wilkinson made key 
suggestions regarding the design of the 
correlation radiometer. Jeff Peterson 
reviewed the manuscript 
and made helpful suggestions. 
Conversations with Robert 
Brandenberger, Josh Gundersen, 
 Ka Lok Ng, Lucio Piccirillo, 
Uros Seljak, and 
 Matias Zaldarriaga have 
refined the focus of the project. 
  This work is supported by NSF grant AST-9318785, a Ford 
Motor Company University 
Research Grant, 
a NASA GSRP Fellowship for BK, 
and a UTRA grant from Brown University for JS.

\newpage
\figcaption{Schematic Dependence of CMB Polarization on Angular Scale and Recombination Scenario. 
In non-standard models which predict a prolonged recombination, the angular scale and level of  
polarization are larger than for standard models.
\label{figthy1}}

\figcaption{Polarization Power Spectra and Window function for POLAR. Three spectra are displayed for 
a CDM ($\Omega = 1, \Omega_B = 0.05, H = 65$,) universe with three different redshifts of reionization 
($z_{{\rm ri}} = 0, 50, 100$). Also shown (solid line) is the single beam window function of POLAR. 
\label{figthy2}}

\figcaption{POLAR Schematic. The correlation polarimeter is a wideband, total-power device. An 
orthomode transducer separates the two orthogonal linear polarization components of the incident 
radiation. HEMT amplifiers are cooled to 15K inside of a dewar. The remaining components are at 
ambient temperature. \label{figexpt1}}

\figcaption{Polarized Foreground Spectra at Millimeter Wavelengths. Spectra of expected polarized 
radiation sources at high galactic latitudes, are shown for a 7$\arcdeg$ beam. A 3 $\mu$K polarized CMB 
signal is shown, corresponding to 10\% of the 10$^{-5}$ CMB anisotropy. At frequencies lower than 90 
GHz the polarization signal is dominated by galactic synchrotron emission (up to 75\% polarized, as 
shown). Galactic bremsstrahlung radiation is not polarized in direct emission, but can be up to 10\% 
polarized (as shown) after Thomson scattering. Galactic dust is shown conservatively with 100\% 
polarization. \label{figforg1}}

\figcaption{Recovery of CMB Polarization Intensity Coefficient vs.  System
Noise. This plot shows the results of the simulations described in section 5.
The vertical axis is the error in the recovered CMB polarization intensity for a
single pixel, in $\mu$K.  The plotted symbols are the Monte Carlo simulation
results for two different values of the synchrotron spectral index; the solid
line is the analytic result for the Foreground Degradation Factor. The analytic
and Monte Carlo results agree extremely well.  This figure demonstrates the
feasibility of detecting a CMB signal polarized at the $\sim 1-3\mu$K level
using a two-frequency configuration.  For such a detection the system noise,
defined over the entire observing time, must be below 1$\mu$K.
\label{figsub1}}

\figcaption{Recovery of Polarization Orientation Angle vs. System Noise. This
plot shows the results of the simulation described in section 5. The vertical
axis is the error in the recovered CMB polar angle for a single pixel, in
degrees. The plotted symbols are the Monte Carlo simulation results for two
different values of the synchrotron spectral index. Similar to the polarized
intensity recovery figure 5, accuracy in the recovery of the CMB
polar angle depends primarily on the system noise, which is defined over the
entire observing time. Recovery of the CMB polar angle with errors less than
$30\arcdeg$ is only possible if the system noise is lower than $1\mu$K.
\label{figsub2}}

\figcaption{Expected rms polarization in reionized universes. The solid line is the magnitude of the rms 
CMB polarization as a function of redshift of reionization. All reionization scenarios assume total 
ionization ($x=1$), except for $z_{{\rm ri}} = 0$, which represents no reionization. The gray band 
represents the $1\sigma$ experimental uncertainty for POLAR, observing 36 pixels for a total time of 
$1\times 10^7$ seconds, with  NET$= 460 \mu$K$s^{\frac{1}{2}}$, after simulated foreground 
subtraction.
\label{figobs1}}

\clearpage
\begin{deluxetable}{lccr}
\footnotesize
\tablewidth{430.69522pt}
\tablenum{1}
\tablewidth{0pt}
\tablecaption{Experimental Limits on Linear Polarization (95\% Confidence Level)}
\tablehead{
\colhead{Reference}           & \colhead{Frequency (GHz)} &
\colhead{Sky Coverage}          & \colhead{Limit $T_{{\rm pol}}/T_{{\rm cmb}}$}  }

\startdata
Penzias \& Wilson 1965 & 4.0 & scattered & 0.1 \nl
Caderni {\em et al.} 1978 & 100$-$600 & near galactic center &  0.001$-$0.01 \nl
Nanos 1979 & 9.3 & declination$= +40\arcdeg$ & $6\times 10^{-4}$ \nl
Lubin \& Smoot 1979 & 33 & delinations 38\arcdeg, 53\arcdeg, 63\arcdeg & $3\times 10^{-4}$\nl
Lubin \& Smoot 1981 & 33 & 11 declinations -37\arcdeg to +63\arcdeg & $6\times 10^{-5}$\nl
Partridge {\em et al.} 1988 & 5 & $43\arcmin \times 43\arcmin$ region, declination 80\arcdeg &$ 4 \times 10^{-5}$\nl
Wollack {\em et al.} 1993 & $26 - 36$ & Degree scales, about NCP & $9\times 10^{-6}$\nl
Netterfield {\em et al.} 1995 & $26 - 46$ & Degree scales, about NCP & $6\times 10^{-6}$\nl
\enddata
\end{deluxetable}

\clearpage
\begin{deluxetable}{lcr}
\footnotesize
\tablewidth{430.69522pt}
\tablenum{2}
\tablewidth{0pt}
\tablecaption{Expected Systematic Effects}
\tablehead{
\colhead{Effect}          &
\colhead{Origin}    &\colhead{Removal/Control Method}  }
\startdata

Mechanical Strain\tablenotemark{a}  & Instrument Rotation & Vertical Drift Scan \nl
Magnetic Coupling\tablenotemark{b}  & Rotation in Earth's Field &  No Ferrite Components\nl
Microphonics\tablenotemark{c}  & Mechanical Vibration &  Mechanical Isolation \nl
Electromagnetic Interference\tablenotemark{d} & Local Sources & Shield/Filter\nl
Offsets\tablenotemark{e} &Polarization Cross-Coupling&Isolation/Instrument Rotation\nl
Radio-Frequency Interference\tablenotemark{f} & In Band Sources & Identification and Data Editing\nl
Thermal Variations\tablenotemark{g} & Diurnal/Environment & Temp Control/Shielding \nl
Sidelobe Pickup\tablenotemark{h} & Sun/Moon/Earth & Low Sidelobe Antenna/Shielding
\enddata

\tablenotetext{a}{\scriptsize A problem with any radiometer that must move in a
gravitational field is position-dependent stress and strain on waveguide joints,
etc.  In POLAR these problems are minimized by staring at the zenith, so no
gravitational torques are present.  The rotation speed is slow, $\sim 1$ rpm so
accelerations on stopping and starting rotation are small. }

\tablenotetext{b}{\scriptsize A particular concern is the coupling of the Earth's magnetic
field to the radiometer.  The {\em COBE} DMR had ferrite Dicke switches which
produced a spurious signal at the $\simeq$0.1 mK level (Kogut et al. 1996b).
POLAR has no ferrite components, but other components such as amplifiers, etc.,
may have a low-level magnetic field dependence. Modulation of these effects can
be minimized by maintaining a constant orientation of rotation axis with respect
to the Earth's field}

\tablenotetext{c}{\scriptsize The effects of vibrations that occur during rotation are
reduced by taking data while the instrument is stationary and by stiffening the
support structures.}

\tablenotetext{d}{\scriptsize This effect can be controlled by Faraday shielding the
instrument and by filtering electrical lines.}

\tablenotetext{e}{\scriptsize Cross polarization in the antenna and the OMT create
correlated signals in both arms of the polarimeter, which in turn produce an
offset signal. If this offset is stable on the timescale of instrument rotation
it is subtracted by phase-sensitive detection at the rotation frequency.}

\tablenotetext{f}{\scriptsize RF sources that occur in the radiometer RF band or IF band are
becoming increasingly troublesome.  Of particular concern in the future will be
communications satellites operating in the bands of interest.}

\tablenotetext{g}{\scriptsize Temperature variations in the radiometer that occur at the
rotation rate of the instrument can be mitigated by active temperature control
and by shielding the instrument from the Sun. The latter function is naturally
performed by the ground shields so that the antenna and receiver are completely
shielded.}

\tablenotetext{h}{\scriptsize The polarimeter must be able to reject or discriminate against
emission from the Sun, Moon, and Earth, which appear only in the sidelobes of
the beam. None of these sources are expected to be significantly polarized, but
asymmetry in the antenna response to the two linear polarizations will create
spurious signals. Requiring the total power from these sources to lie below 1
$\mu$K demands 75 dB and 63 dB sidelobe rejection for the Sun and Moon,
respectively. Assuming 20 dB rejection from the ground screen, this level of
rejection can be achieved with the corrugated horn antenna (Janssen et al. 1979)
if data are rejected when these sources lie closer to the zenith than $50
\arcdeg$ and $30 \arcdeg$ respectively. Binning of the data in Sun-centered or
Moon-centered coordinates will allow us to uncover correlations between the
position of these objects and the response of the polarimeter.}

\end{deluxetable}


\newpage
\begin{references}
\reference{basko} Basko, M. M., \& Polnarev, A.G. 1980, \mnras, 191, 207
\reference{ben1} Bennett, C. L., et al. 1992, \apj, 396, L7
\reference{ben} Bennett, C. L., et al.  1993, \apj, 414, L77
\reference{ben2} Bennett, C. L., et al.  1996, \apj, 464, L1
\reference{bond} Bond, J., \& Efstathiou, G.  1984, \apj, 285, L45
\reference{bond} Bond, J., \& Efstathiou , G. 1987 \mnras, 226, 655
\reference{bran} Brandt, W., et al. 1994, \apj,  424, 1
\reference{brouw} Brouw, W. N., \&  Spoelstra, T.A. 1976, \aaps, 26, 129
\reference{caderni} Caderni, N. 1978, \prd, 17, 1908
\reference{chandra} Chandrasekhar, S. 1960, Radiative Transfer, Dover, New York
\reference{cobra} Visit the Planck Surveyor Homepage: http://astro.estec.esa.nl/SA-general/Projects/Cobras/cobras.html
\reference{cortiglioni} Cortiglioni, S., \& Spoelstra, T. 1995, Astronomy and Astrophysics, 302, 1
\reference{coul} Coulson, D., Crittenden, R. G., \& Turok, N. G. 1994, \prl, 73, 2390
\reference{crill} Crill, B. P.  1995, personal communication
\reference{crit2} Crittenden, R. G., Davis, R., \& Steinhardt, P. 1993 , \apj, 417, L13
\reference{crit}Crittenden, R., Coulson, D., \& Turok, N. 1995, \prd, 52, r5402
\reference{dod} Dodelson, S. 1995, astro-ph 9512021, submitted \apj
\reference{dur} Durrer, R. 1993, astro-ph/9311039 
\reference{fixsen} Fixsen, D., et al. 1996, \apj, 473, 576
\reference{franc} Franceschini, A., et al. 1989, \apj, 344, 35
\reference{frew}Frewin, R., Polnarev, A., \& Coles, P. 1994, \mnras, 266, L21
\reference{fuj} Fujimoto, K., 1964, IEEE-MTT , March p. 203
\reference{good} Gooding, A. K., et al. \apj, 1991, 372, l5
\reference{gunn} Gunn, J. E.. \& Peterson, B. A. 1965, \apj, 142, 1633
\reference{har1} Harari, D. D.. \& Zaldarriaga, M. 1993, Physics Letters B, 319, 96
\reference{har2}  Harari, D. D. , Hayward, J. D., \& Zaldarriaga, M. 1996, \prd, 55, 1841
\reference{hu} Hu, W. 1995, PhD Thesis U.C. Berkeley
\reference{janssen} Janssen, M.A.. et al. 1979, IEEE Trans. Antennas and Prop. 
27(4), 551
\reference{kamionkowski} Kamionkowski, M., et al. 1996, astro-ph 9609132
\reference{keating} Keating, B., \& Polnarev, A. 1997, submitted \apj
\reference{kog1} Kogut, A., et al. 1996a, \apj , 460, 1
\reference{kog2} Kogut, A., et al. 1996b, \apj, 470, 653
\reference{kos} Kosowsky, A. 1996, Annals of Physics, 246, 49
\reference{kos1} Kosowsky, A., \& Loeb, A. 1996, \apj, 469, 1
\reference{krauss} Krauss, J. 1982 Radio Astronomy, McGraw-Hill, New York
\reference{liebe}Liebe, H. 1981, Radio Science, 16 1183
\reference{lube} Lubin, P. 1980, Ph.D. Thesis, U.C. Berkeley 
\reference{lub} Lubin, P., \& Smoot, G. 1981, \apj, 245, 1
\reference{lub2}Lubin, P., Melese, P., Smoot, G. 1983, \apj, 273, L51
\reference{map} Visit the MAP Home Page: http://map.gsfc.nasa.gov
\reference{nanos} Nanos, G., 1979, \apj, 232, 341
\reference{nasel}Nasel'ski, P., \& Polnarev, A. 1987, Astrofizika, 26, 327
\reference{neg} Negroponte, J., \& Silk, J. 1980, \prl, 44, 1433
\reference{net} Netterfield, C.B., et al.,  1995, \apj, 474, L69
\reference{ng93} Ng, K. L., \& Ng, K. W. 1995, \prd, 51, 364
\reference{ng3} Ng, K. L., \& Ng, K. W. 1996, \apj, 456, 413
\reference{oz} Ozernoi, L. M., \& Chernomordik, V. V. 1975, Sov. Astron. 20, 260
\reference{part}Partridge, R., et al. 1988, Nature, 331, 146
\reference{peebles} Peebles, P., 1993, Principles of Physical Cosmology, (Princeton: Princeton University Press) 
\reference{pen}Penzias, A., \& Wilson, R. 1965, \apj, 142, 419
\reference{pol85} Polnarev, A. G. 1985, \azh, 62, 1041
\reference{posa} Pospisszalski, M. 1992, IEEE MTT-S Digest, 1369
\reference{posb} Pospisszalski, M. 1995, IEEE MTT-S Digest, 1121
\reference{rees} Rees, M. J. 1968, \apj , 153, L1
\reference{rohlfs} Rohlfs, K. 1990, Tools of Radio Astronomy, (New York: Springer Verlag)
\reference{rosa} Rosencranz, P.W. 1994, personal communication
\reference{rose} Rosencranz, P. W., \&  Staelin, D. H.  1988, Radio Science, 25(5), 721
\reference{ryb} Rybicki, G. B.,\& Lightman, A. 1979, Radiative Processes in Astrophysics, (New York:Wiley)
\reference{sachs} Sachs, R. K., \& Wolfe, A. M. 1967, \apj, 147, 73
\reference{sakia} Sakia, D., \& Salter, C. 1988, \araa, 26, 93
\reference{scott} Scott, D., Silk, J., \& White, W. 1995, Science, 268, 829
\reference{seljak}Seljak, U., \& Zaldarriaga, M. 1996a, \apj, 469, 437
\reference{seljak2} Seljak, U., \& Zaldarriaga, M. 1996b, astro-ph 9609169
\reference{tegmark} Tegmark, M., \& Silk, J, 1993, `New Constraints on
Reionization fron the Compton y-parameter', in Sanz et al., eds, Present and
Future of the CMB, Proceedings of the Workshop in Santander, Spain 28 June - 1
July 1993, (Berlin: Springer-Verlag) 
\reference{tol} Tolman, B. W., 1985, \apj, 290, 1
\reference{whitesred} White, M., \& Srednicki, M. , 1995, \apj, 443, 6
\reference{wilkinson}Wilkinson, D. T. 1995, `A Warning Label for Cosmic
Microwave Background Anisotropy Experiments', in Astbury et al. , eds., Particle
Physics and Cosmology, Proceedings of the Ninth Lake Louise Winter Institute,
p. 110, (Singapore: World Scientific)
\reference{woll} Wollack, E.J., et al., 1993, \apj, 419, L49
\reference{woll2} Wollack, E.J., et al., 1997, \apj, 476, 440
\reference{wri} Wright, E. L. 1987, \apj, 320, 818
\reference{wri1} Wright, E. L., et al., 1991, \apj, 381, 200
\reference{zal} Zaldarriaga, M., \& Harari, D. 1995, \prd, 52, 3276
\reference{zalsolo} Zaldarriaga, M. 1997, \prd, 55, 1822
\reference{zalsel} Zaldarriaga, M., \& Seljak, U. 1997, \prd, 55, 1830 
\end{references}
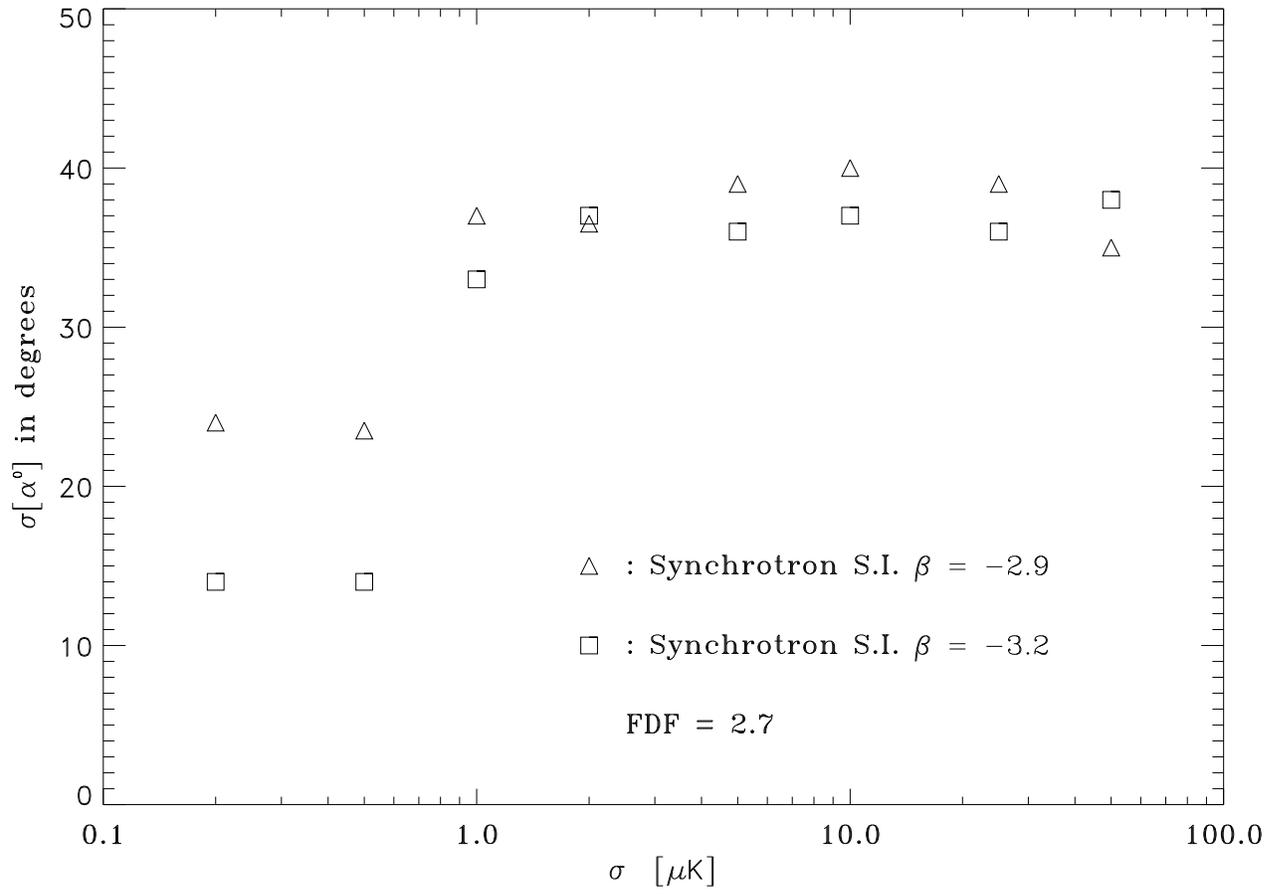
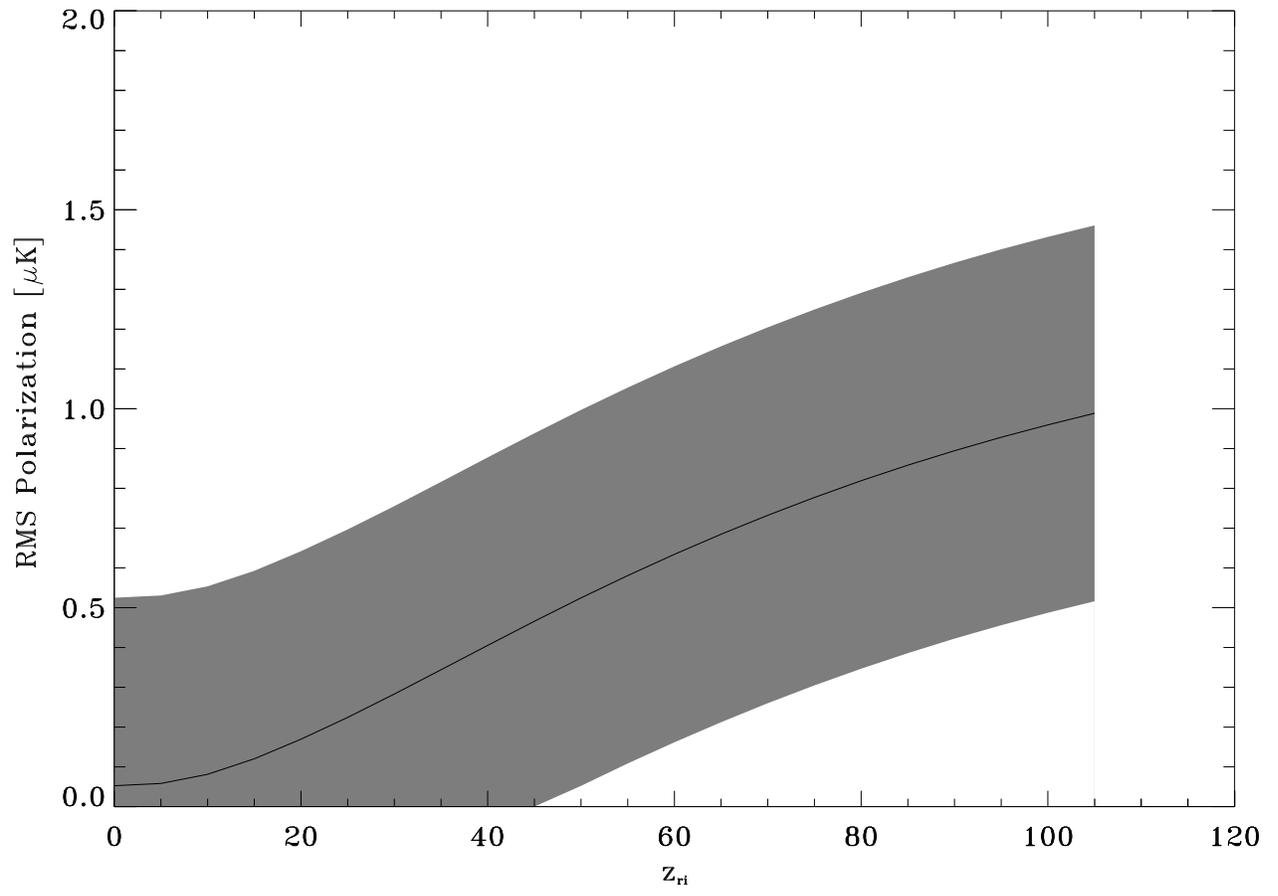
\end{document}